\documentclass[prc,aps,twocolumn,showpacs,amssymb,superscriptaddress,float]{revtex4}

\usepackage{amsmath}
\usepackage{graphicx}
\usepackage{dcolumn}
\usepackage{float}

\begin{document}

\title{Study of neutron-rich calcium isotopes with a realistic shell-model
  interaction}

\author{L. Coraggio}
\affiliation{Istituto Nazionale di Fisica Nucleare, \\
Complesso Universitario di Monte  S. Angelo, Via Cintia - I-80126 Napoli,
Italy}
\author{A. Covello}
\affiliation{Istituto Nazionale di Fisica Nucleare, \\
Complesso Universitario di Monte  S. Angelo, Via Cintia - I-80126 Napoli,
Italy}
\affiliation{Dipartimento di Scienze Fisiche, Universit\`a
di Napoli Federico II, \\
Complesso Universitario di Monte  S. Angelo, Via Cintia - I-80126 Napoli,
Italy}
\author{A. Gargano}
\affiliation{Istituto Nazionale di Fisica Nucleare, \\
Complesso Universitario di Monte  S. Angelo, Via Cintia - I-80126 Napoli,
Italy}
\author{N. Itaco}
\affiliation{Istituto Nazionale di Fisica Nucleare, \\
Complesso Universitario di Monte  S. Angelo, Via Cintia - I-80126 Napoli,
Italy}
\affiliation{Dipartimento di Scienze Fisiche, Universit\`a
di Napoli Federico II, \\
Complesso Universitario di Monte  S. Angelo, Via Cintia - I-80126 Napoli,
Italy}

\date{\today}

\begin{abstract}
We have studied neutron-rich calcium isotopes in terms of the shell
model employing a realistic effective interaction derived from the
CD-Bonn nucleon-nucleon potential. 
The short-range repulsion of the potential is renormalized by way of
the $V_{\rm low-k}$ approach. 
The calculated results are in very good agreement with the available
experimental data, thus supporting our predictions for the hitherto
unknown spectra of $^{53-56}$Ca nuclei. 
In this context, the possible existence of an $N$=34 shell closure
is discussed.
\end{abstract} 


\maketitle

\section{Introduction}
Calcium isotopes with mass number $A>48$ are currently the subject
of great experimental and theoretical interest. 
With an $N/Z$ ratio  $> 1.4$ they lie far from the stability valley
and provide good opportunity to explore the evolution of shell
structure when approaching the neutron drip line. 
In this context, it should be mentioned that the appearance of $N=34$
as a neutron magic number for unstable nuclei was predicted more than
three decades ago within the framework of the energy density formalism
\cite{Beiner75}. 
Recently, some shell-model calculations \cite{Honma04,Honma05} have
revived this issue indicating the existence of a large shell gap at
$N=34$. 
As a consequence, a large number of experiments have been performed in
the region of neutron-rich nuclei around $^{48}$Ca
\cite{Liddick04a,Liddick04b,Dinca05,Gade06,Perrot06,Rejmund07,Mantica08,Fornal08,Maierbeck09,Bhattacharyya09},
aiming at obtaining information about the evolution of the
single-particle (SP) orbitals. 

The experimental data on Ti and Cr isotopes do not evidence such a
shell closure \cite{Dinca05,Fornal05,Zhu06,Zhu07}, even if it should
be noted that the proton-neutron interaction, due to the presence of
valence protons in the $pf$-shell, may hide the neutron shell
structure. 
Consequently, a major experimental issue is currently the
spectroscopic study of $^{53,54}$Ca, since it may provide direct and
unambigous information about the existence of an $N$=34 shell
closure. 

The microscopic theoretical tool to describe the spectroscopic
properties of these nuclei is the shell model, which in recent years
has been extensively used with different two-body effective
interactions. 
The most frequently employed interactions are the KB3G \cite{Poves01},
FPD6 \cite{Richter91}, GXPF1 \cite{Honma02}, and a new version of the
latter dubbed GXPF1A \cite{Honma05}.
The KB3G potential is a monopole-corrected version of the realistic
Kuo-Brown interaction derived some 40 years ago for the $pf$-shell
nuclei \cite{Kuo68a}, while the FPD6 potential has been derived
starting from a modified surface one-boson-exchange potential and then
fitting 61 energy data from 12 nuclei in the mass range 41 to 49
\cite{Richter91}.
The GXPF1 potential has been obtained starting from a $G$-matrix
interaction based on the Bonn C nucleon-nucleon ($NN$) potential
\cite{Hjorth95} and then modifying the two-body matrix elements (TBME)
to fit about 700 energy data in the mass range $A=47-66$ \cite{Honma02}.
These TBME have been recently slightly modified to improve the
description of neutron-rich $pf$-shell nuclei \cite{Honma05}, thus
leading to the GXPF1A interaction.

The above semi-empirical approaches to the derivation of the effective
shell-model interaction have been considered a necessary remedy for
the deficiencies of the original realistic interactions
\cite{Martinez-Pinedo97}.
During the past few years, however, significant progress has been made
in the derivation of realistic shell-model effective interactions. 
This is based on the advent of a new generation of high-precision $NN$
potentials and on a new way to renormalize their strong short-range
repulsion, the $V_{\rm low-k}$ approach \cite{Bogner01,Bogner02},
which has proved to be an advantageous alternative to the traditional
Brueckner $G$-matrix method. 
This has led to the derivation of effective interactions capable of
describing with remarkable accuracy the spectroscopic properties of
nuclei in various mass regions \cite{Coraggio07b,Coraggio07c,Coraggio09}.

On these grounds, we have found it challenging to perform a
shell-model study of the Ca isotopic chain employing an effective
interaction derived from the CD-Bonn potential renormalized by way of
the $V_{\rm low-k}$ approach. 

The aim of the present study is twofold. 
First, to ascertain if our effective interaction does not suffer from
the deficiences that have plagued previous realistic effective
interactions. 
Second, if the answer is in the affirmative, to investigate the shell
structure of neutron-rich Ca isotopes.

The paper is organized as follows. 
In Sec. II we give a brief outline of our calculations and some
details about the choice of the SP energies. 
Sec. III is devoted to the presentation of the results for the calcium
isotopes beyond $^{48}$Ca up to $A$=56.
In Sec. IV we discuss our results, focusing attention on the $N$=34
isotope, and make some concluding remarks. 
In the Appendix the calculated TBME, the employed SP energies, and the
effective single-particle matrix elements of the quadrupole operator
$E2$ are reported.

\section{Outline of calculations}
Our shell-model effective interaction has been derived within the
framework of the perturbation theory \cite{Coraggio09} starting, as
mentioned in the Introduction, from the CD-Bonn $NN$ potential
\cite{Machleidt01b}.
More explicitly, we have first renormalized the high-momentum repulsive
components of the bare $NN$ potential by way of the so-called $V_{\rm
  low-k}$ approach \cite{Bogner01,Bogner02}, which provides a smooth
potential preserving exactly the onshell properties of the original
$NN$ potential.
Next, we have derived the TBME using the well-known $\hat{Q}$-box plus
folded-diagram method \cite{Coraggio09}, where the $\hat{Q}$-box is a
collection of irreducible valence-linked Goldstone diagrams which we
have calculated through third order in the $V_{\rm low-k}$.

The effective interaction $V_{\rm eff}$ can be written in an operator
form as 

\begin{equation}
V_{\rm eff} = \hat{Q} - \hat{Q'} \int \hat{Q} + \hat{Q'} \int \hat{Q} \int
\hat{Q} - \hat{Q'} \int \hat{Q} \int \hat{Q} \int \hat{Q} + ~...~~,
\end{equation}

\noindent
where the integral sign represents a generalized folding operation, 
and $\hat{Q'}$ is obtained from $\hat{Q}$ by removing terms of first
order in $V_{\rm low-k}$.
The folded-diagram series is summed up to all orders using the
Lee-Suzuki iteration method \cite{Suzuki80}.

The model space we have chosen to derive our shell-model effective
interaction is spanned by the four neutron SP levels $0f_{7/2}$,
$0f_{5/2}$, $1p_{3/2}$, and $1p_{1/2}$, located above the
doubly-closed $^{40}$Ca core.
The same choice has been performed in most of the studies on
neutron-rich Ca isotopes, and therefore it allows a direct comparison
between our results and those of other shell-model calculations.

As regards the choice of the neutron SP energies, we have determined
them by reproducing the observed energies of the $(\frac{3}{2}^-)_1$
state in $^{47}$Ca and of the $(\frac{1}{2}^-)_1,~(\frac{5}{2}^-)_2$
states in $^{49}$Ca, whose SP nature is evidenced by the experimental
spectroscopic factors \cite{nndc}.
Our adopted values are $\epsilon_{3/2}-\epsilon_{7/2}=2.7$ MeV,
$\epsilon_{1/2}-\epsilon_{7/2}=5.5$ MeV, and
$\epsilon_{5/2}-\epsilon_{7/2}=8.5$ MeV.
The absolute energy of the neutron $0f_{7/2}$ orbital was placed at
-8.2 MeV, in order to reproduce the experimental binding energy of
$^{49}$Ca with respect to $^{40}$Ca.

The TBME and SP energies used in the present calculation can be found
in the Appendix where, for the sake of completeness, also the
proton-proton and proton-neutron TBME are reported.
It should be pointed out that for protons the Coulomb force has been
explicitly added to the $V_{\rm low-k}$ before constructing $V_{\rm
  eff}$. 

\section{Results}
We have performed shell-model calculations, using the Oslo shell-model
code \cite{EngelandSMC}, for the heavy-mass calcium isotopes from
$^{49}$Ca up to $^{56}$Ca.

\begin{table}[H]
\caption{Experimental negative-parity energy levels (in MeV) of
  $^{49}$Ca \cite{nndc} up to 5.5 MeV excitation energy, compared with
  the calculated ones. The values in parenthesis are the one-neutron
  pickup spectroscopic factors \cite{Uozumi94}.}
\begin{ruledtabular}
\begin{tabular}{ccc}
$J^{\pi}$ & Expt. & Calc. \\
\colrule
$\frac{3}{2}^-$  & 0.000 (0.84) & 0.000 (0.86) \\
$\frac{1}{2}^-$  & 2.023 (0.91) & 2.029 (0.91) \\
$\frac{7}{2}^-$  & ~            & 3.158 \\
$\frac{5}{2}^-$  & 3.585        & 3.300 \\
$\frac{7}{2}^-$  & ~            & 3.731 \\
$\frac{5}{2}^-$  & 3.991 (0.84) & 4.073 (0.86) \\
$\frac{3}{2}^-$  & 4.072        & 3.790 \\
$\frac{1}{2}^-$  & 4.261        &   ~   \\
$\frac{1}{2}^-$  & 4.272        &   ~   \\
$\frac{9}{2}^-$  & ~            & 4.489 \\
$\frac{11}{2}^-$ & ~            & 4.898 \\
$\frac{1}{2}^-$  & 5.444        &   ~   \\
$\frac{5}{2}^-$  & ~            & 5.608 \\
$\frac{3}{2}^-$  & 5.539        & 5.730 \\
\end{tabular}
\end{ruledtabular}
\label{49Catable}
\end{table}

In Table \ref{49Catable} the experimental and calculated
negative-parity levels of $^{49}$Ca are reported up to 5.5 MeV
excitation energy.
It can been seen that, as regards the $(\frac{1}{2}^-)_1$ and
$(\frac{5}{2}^-)_2$ states which we have used to determine the SP
energies, the calculated one-neutron pickup spectroscopic factors
are in very good agreement with experiment, as is the case for the
$\frac{3}{2}^-$ ground state.
From Table \ref{49Catable} it also appears that the calculated
energies of the $(\frac{5}{2}^-)_1$ and $(\frac{3}{2}^-)_{2,3}$ states
are in a good agreement with the experimental values, while above the
yrast state the theory does not predict any other $\frac{1}{2}^-$ state
below 7.5 MeV.

\begin{table}[H]
\caption{Observed positive-parity energy levels (in MeV) of
  $^{50}$Ca \cite{Rejmund07,nndc} up to 5.2 MeV excitation energy,
  compared with the calculated ones. The energies of the states whose
  parity and/or angular momentum are uncertain are reported in
  parenthesis.}
\begin{ruledtabular}
\begin{tabular}{ccc}
$J^{\pi}$ & Expt. & Calc. \\
\colrule
$0^+$  & 0.000   & 0.000 \\
$2^+$  & 1.026   & 0.953 \\
$2^+$  & (3.004) & 2.941 \\
$1^+$  & (3.532) & 3.547 \\
$2^+$  & (4.036) & 3.582 \\
$0^+$  & (4.470) & 4.926 \\
$4^+$  & 4.515   & 4.567 \\
$3^+$  & ~       & 4.645 \\
$4^+$  & ~       & 4.798 \\
$5^+$  & ~       & 5.131 \\
$1^+$  & ~       & 5.167 \\
$2^+$  & (4.870) & 5.231 \\
\end{tabular}
\end{ruledtabular}
\label{50Catable}
\end{table}

The calculated and experimental \cite{Rejmund07} positive parity
states of $^{50}$Ca are reported in Table \ref{50Catable}, we see that
our calculations reproduce quite well the observed energies.
As regards the structure of the states, we find that the four
lowest excited $2^+$ states are dominated by the configurations $\nu
(0f_{7/2})^8 (1p_{3/2})^2$, $\nu (0f_{7/2})^8 (1p_{3/2})^1
(1p_{1/2})^1$, $\nu (0f_{7/2})^7 (1p_{3/2})^3$, $\nu (0f_{7/2})^8
(0f_{5/2})^1 (1p_{3/2})^1$, respectively.
The $\nu (0f_{7/2})^8 (1p_{3/2})^1 (1p_{1/2})^1$ configuration
provides also the structure of the first excited $1^+$ state, 
which is predicted to lie above the second $2^+$ state, in agreement
with what is observed.
This, as mentioned in Refs. \cite{Rejmund07,Casten90}, may be related
to the larger spatial overlap between $1p_{3/2}$ and $1p_{1/2}$ orbits
in the $2^+$ state relative to the $1^+$ state as due to the
requirement of antisymmetrization.

We have also calculated the $B(E2;2^+_1 \rightarrow 0^+_1)$ transition
rates employing an effective operator obtained at third order in
perturbation theory, consistently with the derivation of $H_{\rm
  eff}$.
The calculated value is $10.9~{\rm e^2 fm^4}$ to be compared with the
value $7.5 \pm 0.2~ {\rm e^2 fm^4}$ obtained in a recent experiment
\cite{Valiente-Dobon09}.
In Tables \ref{tableeffn}, \ref{tableeffp} we report the effective
reduced single-neutron and -proton matrix elements of the $E2$
operator.

Our calculation predicts two $4^+$ states in a very small energy
interval $4.6 \div 4.8$ MeV.
The first one belongs mainly to the $\nu (0f_{7/2})^7 (1p_{3/2})^3$
configuration and may be identified with the observed state at 4.515
MeV excitation energy \cite{Rejmund07}, while the second one
essentially arises from the $\nu (0f_{7/2})^8 (1p_{3/2})^1
(0f_{5/2})^1$ configuration. 

The experimental $(0^+_2)$ state at 4.47 MeV \cite{Bjerregaard67} may
contain significant contributions from $2p-2h$ $Z=20$ cross-shell
excitations.
As a matter of fact, in $^{48}$Ca two excited $0^+$ states are
observed; the first one at 4.28 MeV excitation energy is associated
with the above mentioned $2p-2h$ excitations \cite{Brown98}, while the
second one at 5.46 MeV excitation energy belongs mainly to the $\nu
(0f_{7/2})^6 (1p_{3/2})^2$ configuration.
In this connection, it should be mentioned that our calculated energy
of the latter state is 5.22 MeV.
So, it would be desirable to obtain more experimental information
about the first two $0^+$ excited states also in $^{50}$Ca, in order
to have a better understanding of the role of $Z=20$ cross-shell
excitations.

\begin{table}[H]
\caption{Experimental negative-parity energy levels (in MeV) of
  $^{51}$Ca \cite{Rejmund07} up to 4.5 MeV excitation energy,
  compared with the calculated ones. The energies of the states whose
  parity and/or angular momentum are uncertain are reported in
  parenthesis.}
\begin{ruledtabular}
\begin{tabular}{ccc}
$J^{\pi}$ & Expt. & Calc. \\
\colrule
$\frac{3}{2}^-$  & (0.000) & 0.000 \\
$\frac{1}{2}^-$  & (1.721) & 1.586 \\
$\frac{5}{2}^-$  & (2.379) & 2.269 \\
$\frac{3}{2}^-$  & (2.937) & 2.932 \\
$\frac{7}{2}^-$  & (3.437) & 3.429 \\
$\frac{5}{2}^-$  & (3.479) & 3.552 \\
$\frac{9}{2}^-$  & (4.322) & 4.435 \\
$\frac{1}{2}^-$  & ~       & 4.470 \\
$\frac{7}{2}^-$  & ~       & 4.488 \\
\end{tabular}
\end{ruledtabular}
\label{51Catable}
\end{table}

In Table \ref{51Catable} we have reported experimental and calculated
negative-parity states of $^{51}$Ca.
We see that also for this nucleus the comparison between theory and
experiment is remarkably good, the largest discrepancy being about 100
keV.

As regards our calculated wavefunctions, all low-lying states are
mainly constructed as 8 neutrons in the $0f_{7/2}$ orbital - the
$^{48}$Ca core  - plus the 3 extra neutrons in the remaining SP
levels, except the $(\frac{7}{2}^-)_1$ state whose main configuration
is made up by one neutron-hole in $^{48}$Ca core and 4 neutrons
filling the $1p_{3/2}$ orbital.
More precisely, the $(\frac{3}{2}^-)_1$ ground state is dominated by
the $^{48}$Ca$\otimes (\nu 1p_{3/2})^3 $ configuration, the
$(\frac{1}{2}^-)_1,~(\frac{5}{2}^-)_1,~(\frac{3}{2}^-)_2$ states by
the $^{48}$Ca$\otimes (\nu 1p_{3/2})^2 (\nu 1p_{1/2})^1 $
configuration, and the $(\frac{5}{2}^-)_2,~(\frac{9}{2}^-)_1$ states
by the $^{48}$Ca$\otimes (\nu 1p_{3/2})^2 (\nu 0f_{5/2})^1 $
configuration.

\begin{table}[H]
\caption{Low-lying positive-parity energy levels (in MeV) of
  $^{52}$Ca \cite{Rejmund07,Gade06,nndc} compared with the calculated ones. 
  The energies of the states whose parity and/or angular momentum are
  uncertain are reported in parenthesis.}
\begin{ruledtabular}
\begin{tabular}{ccc}
$J^{\pi}$ & Expt. & Calc. \\
\colrule
$0^+$  & 0.000   & 0.000 \\
$2^+$  & 2.562   & 2.396 \\
$1^+$  & (3.150) & 3.135 \\
$0^+$  & ~       & 4.040 \\
$2^+$  & ~       & 4.432 \\
$3^+$  & ~       & 4.568 \\
$4^+$  & ~       & 4.743 \\
\end{tabular}
\end{ruledtabular}
\label{52Catable}
\end{table}

The nucleus $^{52}$Ca is the last one of this isotopic chain for which
a few experimental excited states have been observed
\cite{Rejmund07,Gade06,Perrot06}.
Its energies are reported in Table \ref{52Catable} together with the
calculated ones up to 5 MeV.
Note that the neutron binding energy for $^{52}$Ca is about 4.7
MeV \cite{Audi03}.

\begin{table}[H]
\caption{Theoretical negative-parity energy levels (in MeV) of
  $^{53}$Ca up to 4.0 MeV excitation energy.}
\begin{ruledtabular}
\begin{tabular}{cc}
$J^{\pi}$ & Calc. \\
\colrule
$\frac{1}{2}^-$  & 0.000 \\
$\frac{5}{2}^-$  & 2.039 \\
$\frac{3}{2}^-$  & 2.402 \\
\end{tabular}
\end{ruledtabular}
\label{53Catable}
\end{table}

The jump in energy of the first excited $2^+$ state, when going from
 $^{50}$Ca to $^{52}$Ca, reflects the subshell filling of the
 $1p_{3/2}$ orbital.
In fact, to construct the $2^+_1$ state in $^{52}$Ca a pair of
neutrons in the $1p_{3/2}$ orbital has to be broken and one neutron
promoted to the $1p_{1/2}$ SP level. 

For the heavier Ca isotopes $^{53,54,55}$Ca, only the ground states
have been identified with spin and parity assignment \cite{Mantica08}.
In Tables \ref{53Catable},\ref{54Catable}, \ref{55Catable} we have
reported our predicted spectra for the three nuclei up to 4.0 MeV
excitation energy.

\begin{table}[H]
\caption{Theoretical positive-parity energy levels (in MeV) of
  $^{54}$Ca up to 4.0 MeV excitation energy.}
\begin{ruledtabular}
\begin{tabular}{cc}
$J^{\pi}$ & Calc. \\
\colrule
$0^+$  & 0.000 \\
$2^+$  & 2.061 \\
$3^+$  & 2.748 \\
$0^+$  & 3.138 \\
\end{tabular}
\end{ruledtabular}
\label{54Catable}
\end{table}

\begin{table}[H]
\caption{Theoretical negative-parity energy levels (in MeV) of
  $^{55}$Ca up to 4.0 MeV excitation energy.}
\begin{ruledtabular}
\begin{tabular}{cc}
$J^{\pi}$ & Calc. \\
\colrule
$\frac{5}{2}^-$  & 0.000 \\
$\frac{1}{2}^-$  & 1.104 \\
$\frac{3}{2}^-$  & 1.541 \\
$\frac{7}{2}^-$  & 1.931 \\
$\frac{5}{2}^-$  & 2.390 \\
$\frac{9}{2}^-$  & 2.914 \\
$\frac{5}{2}^-$  & 3.472 \\
$\frac{3}{2}^-$  & 3.547 \\
$\frac{3}{2}^-$  & 3.938 \\
\end{tabular}
\end{ruledtabular}
\label{55Catable}
\end{table}

\begin{figure}[H]
\begin{center}
\includegraphics[scale=0.4,angle=90]{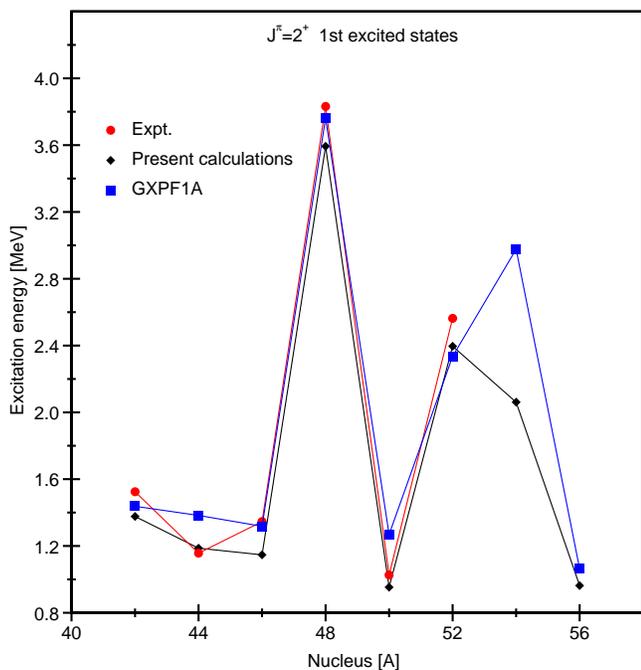}
\caption{(Color online) Experimental \cite{nndc} and calculated
  excitation energies of the yrast $J^{\pi}=2^+$ states for calcium
  isotopes from $A=42$ to 56. Black diamonds refer to present
  calculations, blue squares to those with GXPF1A potential.}
\label{fig2pgx}
\end{center}
\end{figure}

\section{Discussion and concluding remarks}
Shell-model effective interactions derived from the free $NN$ potential
were proposed for the $pf$-shell in Refs. \cite{Kuo68a,Hjorth95}.
It has been shown, however, that while these interactions give a good
description of the spectroscopy of few-valence-nucleon systems, they
fail for many-valence-particle nuclei. 
This is evidenced by the poor description of the closure properties of
$^{48}$Ca and  $^{56}$Ni \cite{McGrory70,Hjorth95}, and justifies the
use of empirically modified versions of these interactions in this
mass region.

As mentioned in the Introduction, our realistic effective interaction
is based on recent progress achieved in its derivation, and the good
agreement of our results with the available experimental data seems to
testify to its reliability.

\begin{figure}[H]
\begin{center}
\includegraphics[scale=0.4,angle=0]{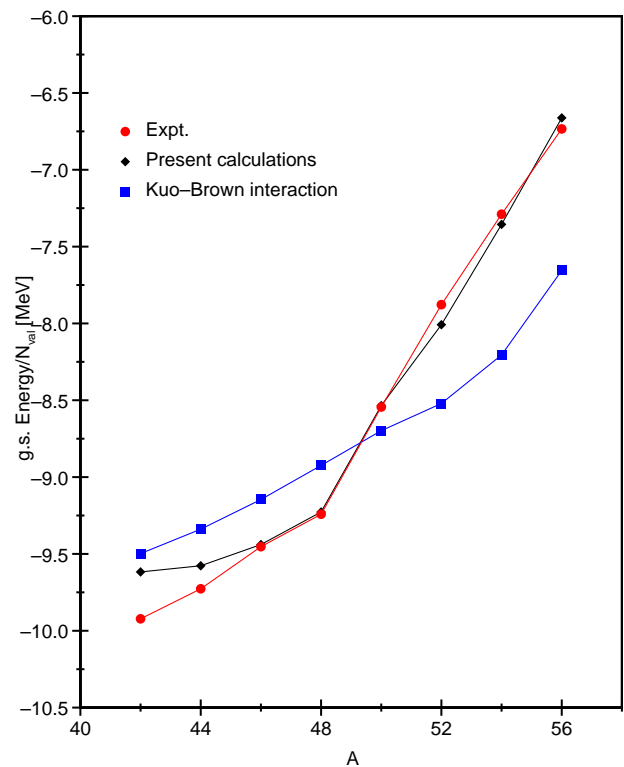}
\caption{(Color online)  Experimental \cite{Audi03} and calculated
  ground-state  energies per valence neutron for calcium isotopes from
  $A=42$ to 56. $N_{\rm val}$ is the number of valence neutrons. Black
  diamonds refer to present calculations, blue squares to those with
  Kuo-Brown interaction.}
\label{figgse}
\end{center}
\end{figure}

This is further confirmed by the inspection of Fig. \ref{fig2pgx},
where the experimental excitation energies of the yrast $2^+$ state
are reported as a function of $A$, and compared with those calculated
with both our shell-model hamiltonian and the GXPF1A one.
It can be seen that our calculations reproduce nicely the observed
energies of the first excited $2^+$ states without any need of
monopole modification.
The good quality of the $T=1$ monopole part of our interaction is also
evidenced in Fig. \ref{figgse}, where we
have plotted the calculated binding energies per valence neutron as a
function of $A$. 
We see a good agreement with the experimental data along all the
isotopic chain, at variance with the Kuo-Brown interaction, whose
results are reported for comparison.

We have pointed out in the Introduction that an issue of great
interest in the study of heavy-mass calcium isotopes is the possible
appearance of the ``magic number'' $N$=34.
Our results do not provide evidence of this magic number, since they
do not predict an increase of the $2^+_1$ excitation energy in
$^{54}$Ca with respect to $^{52}$Ca.
It should be mentioned that calculations performed with the KB3G
\cite{Caurier05} and FPD6 \cite{Poves01} interactions do not predict a
shell closure at $N$=34, that is instead predicted when using the
GXPF1A interaction \cite{Honma05}.

To understand the different behavior of the two theoretical curves of
Fig. \ref{fig2pgx}, it is useful to consider the effective single
particle energies (ESPE) for calcium isotopes (see for instance
Ref. \cite{Utsuno99}).

We recall here that the ESPE are related to the monopole part of the
shell-model hamiltonian, thus reflecting the angular-momentum-averaged
effects of the two-body interaction for a given nucleus.
The ESPE of a level is defined as the one-neutron separation
energy of this level, and is calculated in terms of the bare
$\epsilon_j$ and the monopole part of the interaction, ${\rm ESPE}(j)
= \epsilon_j + \sum_{j'} V_{j j'} n_{j'}$, where the sum runs on the
model-space levels $j'$, $n_{j}$ being the number of particles in the
level $j$ and $V_{j j'}$ the angular-momentum-averaged interaction
$V_{j j'} = \sum_J (2J +1) \langle j j' | V | j j' \rangle_J / \sum_J
(2J +1)$. 

\begin{figure}[H]
\begin{center}
\includegraphics[scale=0.4,angle=90]{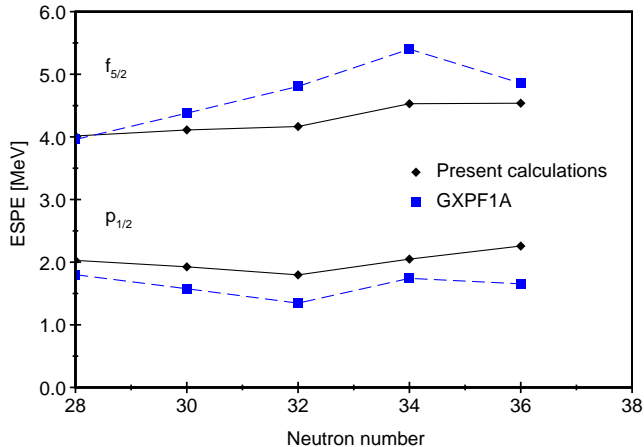}
\caption{(Color online) Effective single-particle energies of the
  neutron $1p_{1/2}$ and $0f_{5/2}$ orbits relative to the
  $1p_{3/2}$ from $N=28$ to 36. Black diamonds refer to present
  calculations, blue squares to those with GXPF1A potential.}
\label{figESPE}
\end{center}
\end{figure}

In Fig. \ref{figESPE} we report the ESPE of the neutron $1p_{1/2}$ and
$0f_{5/2}$ orbits relative to the $1p_{3/2}$ orbit as a function of
the neutron number.
The appearance of the $N$=34 magic number is obviously related to the
existence of a large gap between the $1p_{1/2}$ and $0f_{5/2}$
orbits. 
For $N =$28 the two effective interactions predict almost the same gap
of about 2 MeV, but, when adding neutrons, this gap remains
essentially unchanged up to $N$=34 in our calculations, while the
GXPF1A interaction predicts that the difference between the two ESPE
increases  rapidly up to a value of about 3.6 MeV for $N$=34.
These contrasting scenarios for the shell evolution are related to the 
different $T=1$ monopole properties of the two effective interactions.
In this connection, it should be pointed out that for the
$(0f_{5/2}1p_{3/2})$ configuration the monopole is -0.09 MeV for our
interaction, and 0.12 MeV for GXPF1A.
For the $(0f_{5/2}1p_{1/2})$ configuration, the present interaction
gives a monopole equal to 0.02 MeV, while GXPF1A gives 0.08 MeV.

For the sake of completeness, it should be mentioned that recently new
results have been reported from a one-neutron knockout reaction on
$^{56}$Ti \cite{Maierbeck09}. 
These data establish that the ground state of $^{55}$Ti has $J^\pi =
\frac{1}{2}^-$. 
However, both shell-model predictions using GXPF1A and our interaction
are in agreement with this assignment, showing that it does not help
shedding light on the problem of the $N$=34 shell closure.

We hope that the present work may contribute to the understanding of
the shell evolution and nuclear structure properties when moving
toward the neutron drip line in the Ca region.

\section*{Appendix}
\renewcommand{\thetable}{A.\Roman{table}}
\setcounter{table}{0}

\begin{table}[H]
\caption{Neutron-neutron TBME (in MeV). They are antisymmetrized and
  normalized.}
\begin{ruledtabular}
\begin{tabular}{cccc}
$n_a l_a j_a ~ n_b l_b j_b ~ n_c l_c j_c ~ n_d l_d j_d $ & $J$ & $T_z$
  &  TBME \\
\colrule
 $ 0f_{ 7/2}~ 0f_{ 7/2}~ 0f_{ 7/2}~ 0f_{ 7/2}$ &  0 & -1 & -1.959 \\
 $ 0f_{ 7/2}~ 0f_{ 7/2}~ 0f_{ 5/2}~ 0f_{ 5/2}$ &  0 & -1 & -2.807 \\
 $ 0f_{ 7/2}~ 0f_{ 7/2}~ 1p_{ 3/2}~ 1p_{ 3/2}$ &  0 & -1 & -1.328 \\
 $ 0f_{ 7/2}~ 0f_{ 7/2}~ 1p_{ 1/2}~ 1p_{ 1/2}$ &  0 & -1 & -1.068 \\
 $ 0f_{ 5/2}~ 0f_{ 5/2}~ 0f_{ 5/2}~ 0f_{ 5/2}$ &  0 & -1 & -0.949 \\
 $ 0f_{ 5/2}~ 0f_{ 5/2}~ 1p_{ 3/2}~ 1p_{ 3/2}$ &  0 & -1 & -1.024 \\
 $ 0f_{ 5/2}~ 0f_{ 5/2}~ 1p_{ 1/2}~ 1p_{ 1/2}$ &  0 & -1 & -0.477 \\
 $ 1p_{ 3/2}~ 1p_{ 3/2}~ 1p_{ 3/2}~ 1p_{ 3/2}$ &  0 & -1 & -1.130 \\
 $ 1p_{ 3/2}~ 1p_{ 3/2}~ 1p_{ 1/2}~ 1p_{ 1/2}$ &  0 & -1 & -1.240 \\
 $ 1p_{ 1/2}~ 1p_{ 1/2}~ 1p_{ 1/2}~ 1p_{ 1/2}$ &  0 & -1 & -0.204 \\
 $ 0f_{ 7/2}~ 0f_{ 5/2}~ 0f_{ 7/2}~ 0f_{ 5/2}$ &  1 & -1 &  0.183 \\
 $ 0f_{ 7/2}~ 0f_{ 5/2}~ 0f_{ 5/2}~ 1p_{ 3/2}$ &  1 & -1 & -0.194 \\
 $ 0f_{ 7/2}~ 0f_{ 5/2}~ 1p_{ 3/2}~ 1p_{ 1/2}$ &  1 & -1 & -0.059 \\
 $ 0f_{ 5/2}~ 1p_{ 3/2}~ 0f_{ 5/2}~ 1p_{ 3/2}$ &  1 & -1 &  0.091 \\
 $ 0f_{ 5/2}~ 1p_{ 3/2}~ 1p_{ 3/2}~ 1p_{ 1/2}$ &  1 & -1 & -0.099 \\
 $ 1p_{ 3/2}~ 1p_{ 1/2}~ 1p_{ 3/2}~ 1p_{ 1/2}$ &  1 & -1 &  0.333 \\
 $ 0f_{ 7/2}~ 0f_{ 7/2}~ 0f_{ 7/2}~ 0f_{ 7/2}$ &  2 & -1 & -1.015 \\
 $ 0f_{ 7/2}~ 0f_{ 7/2}~ 0f_{ 7/2}~ 0f_{ 5/2}$ &  2 & -1 &  0.169 \\
 $ 0f_{ 7/2}~ 0f_{ 7/2}~ 0f_{ 7/2}~ 1p_{ 3/2}$ &  2 & -1 & -0.672 \\
 $ 0f_{ 7/2}~ 0f_{ 7/2}~ 0f_{ 5/2}~ 0f_{ 5/2}$ &  2 & -1 & -0.713 \\
 $ 0f_{ 7/2}~ 0f_{ 7/2}~ 0f_{ 5/2}~ 1p_{ 3/2}$ &  2 & -1 &  0.564 \\
 $ 0f_{ 7/2}~ 0f_{ 7/2}~ 0f_{ 5/2}~ 1p_{ 1/2}$ &  2 & -1 & -0.716 \\
 $ 0f_{ 7/2}~ 0f_{ 7/2}~ 1p_{ 3/2}~ 1p_{ 3/2}$ &  2 & -1 & -0.411 \\
\end{tabular}
\end{ruledtabular}
\label{tablenn}
\end{table}

\begin{table}[H]
\begin{ruledtabular}
\begin{tabular}{cccc}
 $ 0f_{ 7/2}~ 0f_{ 7/2}~ 1p_{ 3/2}~ 1p_{ 1/2}$ &  2 & -1 & -0.456 \\
 $ 0f_{ 7/2}~ 0f_{ 5/2}~ 0f_{ 7/2}~ 0f_{ 5/2}$ &  2 & -1 & -0.149 \\
 $ 0f_{ 7/2}~ 0f_{ 5/2}~ 0f_{ 7/2}~ 1p_{ 3/2}$ &  2 & -1 &  0.004 \\
 $ 0f_{ 7/2}~ 0f_{ 5/2}~ 0f_{ 5/2}~ 0f_{ 5/2}$ &  2 & -1 & -0.677 \\
 $ 0f_{ 7/2}~ 0f_{ 5/2}~ 0f_{ 5/2}~ 1p_{ 3/2}$ &  2 & -1 &  0.408 \\
 $ 0f_{ 7/2}~ 0f_{ 5/2}~ 0f_{ 5/2}~ 1p_{ 1/2}$ &  2 & -1 & -0.446 \\
 $ 0f_{ 7/2}~ 0f_{ 5/2}~ 1p_{ 3/2}~ 1p_{ 3/2}$ &  2 & -1 & -0.059 \\
 $ 0f_{ 7/2}~ 0f_{ 5/2}~ 1p_{ 3/2}~ 1p_{ 1/2}$ &  2 & -1 & -0.156 \\
 $ 0f_{ 7/2}~ 1p_{ 3/2}~ 0f_{ 7/2}~ 1p_{ 3/2}$ &  2 & -1 & -1.045 \\
 $ 0f_{ 7/2}~ 1p_{ 3/2}~ 0f_{ 5/2}~ 0f_{ 5/2}$ &  2 & -1 & -0.697 \\
 $ 0f_{ 7/2}~ 1p_{ 3/2}~ 0f_{ 5/2}~ 1p_{ 3/2}$ &  2 & -1 &  0.490 \\
 $ 0f_{ 7/2}~ 1p_{ 3/2}~ 0f_{ 5/2}~ 1p_{ 1/2}$ &  2 & -1 & -1.277 \\
 $ 0f_{ 7/2}~ 1p_{ 3/2}~ 1p_{ 3/2}~ 1p_{ 3/2}$ &  2 & -1 & -0.593 \\
 $ 0f_{ 7/2}~ 1p_{ 3/2}~ 1p_{ 3/2}~ 1p_{ 1/2}$ &  2 & -1 & -0.828 \\
 $ 0f_{ 5/2}~ 0f_{ 5/2}~ 0f_{ 5/2}~ 0f_{ 5/2}$ &  2 & -1 & -0.397 \\
 $ 0f_{ 5/2}~ 0f_{ 5/2}~ 0f_{ 5/2}~ 1p_{ 3/2}$ &  2 & -1 &  0.053 \\
 $ 0f_{ 5/2}~ 0f_{ 5/2}~ 0f_{ 5/2}~ 1p_{ 1/2}$ &  2 & -1 & -0.273 \\
 $ 0f_{ 5/2}~ 0f_{ 5/2}~ 1p_{ 3/2}~ 1p_{ 3/2}$ &  2 & -1 & -0.196 \\
 $ 0f_{ 5/2}~ 0f_{ 5/2}~ 1p_{ 3/2}~ 1p_{ 1/2}$ &  2 & -1 & -0.408 \\
 $ 0f_{ 5/2}~ 1p_{ 3/2}~ 0f_{ 5/2}~ 1p_{ 3/2}$ &  2 & -1 &  0.122 \\
 $ 0f_{ 5/2}~ 1p_{ 3/2}~ 0f_{ 5/2}~ 1p_{ 1/2}$ &  2 & -1 &  0.583 \\
 $ 0f_{ 5/2}~ 1p_{ 3/2}~ 1p_{ 3/2}~ 1p_{ 3/2}$ &  2 & -1 &  0.226 \\
 $ 0f_{ 5/2}~ 1p_{ 3/2}~ 1p_{ 3/2}~ 1p_{ 1/2}$ &  2 & -1 &  0.315 \\
 $ 0f_{ 5/2}~ 1p_{ 1/2}~ 0f_{ 5/2}~ 1p_{ 1/2}$ &  2 & -1 & -0.336 \\
 $ 0f_{ 5/2}~ 1p_{ 1/2}~ 1p_{ 3/2}~ 1p_{ 3/2}$ &  2 & -1 & -0.405 \\
 $ 0f_{ 5/2}~ 1p_{ 1/2}~ 1p_{ 3/2}~ 1p_{ 1/2}$ &  2 & -1 & -0.543 \\
 $ 1p_{ 3/2}~ 1p_{ 3/2}~ 1p_{ 3/2}~ 1p_{ 3/2}$ &  2 & -1 & -0.217 \\
 $ 1p_{ 3/2}~ 1p_{ 3/2}~ 1p_{ 3/2}~ 1p_{ 1/2}$ &  2 & -1 & -0.706 \\
 $ 1p_{ 3/2}~ 1p_{ 1/2}~ 1p_{ 3/2}~ 1p_{ 1/2}$ &  2 & -1 & -0.528 \\
 $ 0f_{ 7/2}~ 0f_{ 5/2}~ 0f_{ 7/2}~ 0f_{ 5/2}$ &  3 & -1 &  0.313 \\
 $ 0f_{ 7/2}~ 0f_{ 5/2}~ 0f_{ 7/2}~ 1p_{ 3/2}$ &  3 & -1 & -0.283 \\
 $ 0f_{ 7/2}~ 0f_{ 5/2}~ 0f_{ 7/2}~ 1p_{ 1/2}$ &  3 & -1 &  0.160 \\
 $ 0f_{ 7/2}~ 0f_{ 5/2}~ 0f_{ 5/2}~ 1p_{ 3/2}$ &  3 & -1 & -0.194 \\
 $ 0f_{ 7/2}~ 0f_{ 5/2}~ 0f_{ 5/2}~ 1p_{ 1/2}$ &  3 & -1 & -0.037 \\
 $ 0f_{ 7/2}~ 1p_{ 3/2}~ 0f_{ 7/2}~ 1p_{ 3/2}$ &  3 & -1 &  0.163 \\
 $ 0f_{ 7/2}~ 1p_{ 3/2}~ 0f_{ 7/2}~ 1p_{ 1/2}$ &  3 & -1 &  0.019 \\
 $ 0f_{ 7/2}~ 1p_{ 3/2}~ 0f_{ 5/2}~ 1p_{ 3/2}$ &  3 & -1 &  0.072 \\
 $ 0f_{ 7/2}~ 1p_{ 3/2}~ 0f_{ 5/2}~ 1p_{ 1/2}$ &  3 & -1 & -0.075 \\
 $ 0f_{ 7/2}~ 1p_{ 1/2}~ 0f_{ 7/2}~ 1p_{ 1/2}$ &  3 & -1 &  0.281 \\
 $ 0f_{ 7/2}~ 1p_{ 1/2}~ 0f_{ 5/2}~ 1p_{ 3/2}$ &  3 & -1 &  0.170 \\
 $ 0f_{ 7/2}~ 1p_{ 1/2}~ 0f_{ 5/2}~ 1p_{ 1/2}$ &  3 & -1 & -0.101 \\
 $ 0f_{ 5/2}~ 1p_{ 3/2}~ 0f_{ 5/2}~ 1p_{ 3/2}$ &  3 & -1 &  0.193 \\
 $ 0f_{ 5/2}~ 1p_{ 3/2}~ 0f_{ 5/2}~ 1p_{ 1/2}$ &  3 & -1 &  0.022 \\
 $ 0f_{ 5/2}~ 1p_{ 1/2}~ 0f_{ 5/2}~ 1p_{ 1/2}$ &  3 & -1 &  0.265 \\
 $ 0f_{ 7/2}~ 0f_{ 7/2}~ 0f_{ 7/2}~ 0f_{ 7/2}$ &  4 & -1 & -0.196 \\
 $ 0f_{ 7/2}~ 0f_{ 7/2}~ 0f_{ 7/2}~ 0f_{ 5/2}$ &  4 & -1 & -0.564 \\
 $ 0f_{ 7/2}~ 0f_{ 7/2}~ 0f_{ 7/2}~ 1p_{ 3/2}$ &  4 & -1 & -0.471 \\
 $ 0f_{ 7/2}~ 0f_{ 7/2}~ 0f_{ 7/2}~ 1p_{ 1/2}$ &  4 & -1 & -0.474 \\
 $ 0f_{ 7/2}~ 0f_{ 7/2}~ 0f_{ 5/2}~ 0f_{ 5/2}$ &  4 & -1 & -0.508 \\
 $ 0f_{ 7/2}~ 0f_{ 7/2}~ 0f_{ 5/2}~ 1p_{ 3/2}$ &  4 & -1 &  0.674 \\
 $ 0f_{ 7/2}~ 0f_{ 5/2}~ 0f_{ 7/2}~ 0f_{ 5/2}$ &  4 & -1 &  0.127 \\
 $ 0f_{ 7/2}~ 0f_{ 5/2}~ 0f_{ 7/2}~ 1p_{ 3/2}$ &  4 & -1 & -0.236 \\
 $ 0f_{ 7/2}~ 0f_{ 5/2}~ 0f_{ 7/2}~ 1p_{ 1/2}$ &  4 & -1 & -0.065 \\
 $ 0f_{ 7/2}~ 0f_{ 5/2}~ 0f_{ 5/2}~ 0f_{ 5/2}$ &  4 & -1 & -0.509 \\
 $ 0f_{ 7/2}~ 0f_{ 5/2}~ 0f_{ 5/2}~ 1p_{ 3/2}$ &  4 & -1 &  0.835 \\
 $ 0f_{ 7/2}~ 1p_{ 3/2}~ 0f_{ 7/2}~ 1p_{ 3/2}$ &  4 & -1 & -0.112 \\
 $ 0f_{ 7/2}~ 1p_{ 3/2}~ 0f_{ 7/2}~ 1p_{ 1/2}$ &  4 & -1 & -0.763 \\
 $ 0f_{ 7/2}~ 1p_{ 3/2}~ 0f_{ 5/2}~ 0f_{ 5/2}$ &  4 & -1 & -0.290 \\
 $ 0f_{ 7/2}~ 1p_{ 3/2}~ 0f_{ 5/2}~ 1p_{ 3/2}$ &  4 & -1 &  0.916 \\
 $ 0f_{ 7/2}~ 1p_{ 1/2}~ 0f_{ 7/2}~ 1p_{ 1/2}$ &  4 & -1 & -0.281 \\
 $ 0f_{ 7/2}~ 1p_{ 1/2}~ 0f_{ 5/2}~ 0f_{ 5/2}$ &  4 & -1 & -0.379 \\
 $ 0f_{ 7/2}~ 1p_{ 1/2}~ 0f_{ 5/2}~ 1p_{ 3/2}$ &  4 & -1 &  0.942 \\
 $ 0f_{ 5/2}~ 0f_{ 5/2}~ 0f_{ 5/2}~ 0f_{ 5/2}$ &  4 & -1 & -0.011 \\
\end{tabular}
\end{ruledtabular}
\end{table}

\begin{table}[H]
\begin{ruledtabular}
\begin{tabular}{cccc}
 $ 0f_{ 5/2}~ 0f_{ 5/2}~ 0f_{ 5/2}~ 1p_{ 3/2}$ &  4 & -1 &  0.232 \\
 $ 0f_{ 5/2}~ 1p_{ 3/2}~ 0f_{ 5/2}~ 1p_{ 3/2}$ &  4 & -1 & -0.489 \\
 $ 0f_{ 7/2}~ 0f_{ 5/2}~ 0f_{ 7/2}~ 0f_{ 5/2}$ &  5 & -1 &  0.440 \\
 $ 0f_{ 7/2}~ 0f_{ 5/2}~ 0f_{ 7/2}~ 1p_{ 3/2}$ &  5 & -1 & -0.243 \\
 $ 0f_{ 7/2}~ 1p_{ 3/2}~ 0f_{ 7/2}~ 1p_{ 3/2}$ &  5 & -1 &  0.657 \\
 $ 0f_{ 7/2}~ 0f_{ 7/2}~ 0f_{ 7/2}~ 0f_{ 7/2}$ &  6 & -1 &  0.211 \\
 $ 0f_{ 7/2}~ 0f_{ 7/2}~ 0f_{ 7/2}~ 0f_{ 5/2}$ &  6 & -1 & -1.205 \\
 $ 0f_{ 7/2}~ 0f_{ 5/2}~ 0f_{ 7/2}~ 0f_{ 5/2}$ &  6 & -1 & -1.246 \\
\end{tabular}
\end{ruledtabular}
\end{table}

\begin{table}[H]
\caption{Same as in Table \ref{tablenn}, but for proton-proton TBME.}
\begin{ruledtabular}
\begin{tabular}{cccc}
$n_a l_a j_a ~ n_b l_b j_b ~ n_c l_c j_c ~ n_d l_d j_d $ & $J$ & $T_z$
  &  TBME \\
\colrule
 $ 0f_{ 7/2}~ 0f_{ 7/2}~ 0f_{ 7/2}~ 0f_{ 7/2}$ &  0 &  1 & -1.727 \\
 $ 0f_{ 7/2}~ 0f_{ 7/2}~ 0f_{ 5/2}~ 0f_{ 5/2}$ &  0 &  1 & -2.584 \\
 $ 0f_{ 7/2}~ 0f_{ 7/2}~ 1p_{ 3/2}~ 1p_{ 3/2}$ &  0 &  1 & -1.348 \\
 $ 0f_{ 7/2}~ 0f_{ 7/2}~ 1p_{ 1/2}~ 1p_{ 1/2}$ &  0 &  1 & -1.058 \\
 $ 0f_{ 5/2}~ 0f_{ 5/2}~ 0f_{ 5/2}~ 0f_{ 5/2}$ &  0 &  1 & -0.719 \\
 $ 0f_{ 5/2}~ 0f_{ 5/2}~ 1p_{ 3/2}~ 1p_{ 3/2}$ &  0 &  1 & -1.066 \\
 $ 0f_{ 5/2}~ 0f_{ 5/2}~ 1p_{ 1/2}~ 1p_{ 1/2}$ &  0 &  1 & -0.515 \\
 $ 1p_{ 3/2}~ 1p_{ 3/2}~ 1p_{ 3/2}~ 1p_{ 3/2}$ &  0 &  1 & -0.635 \\
 $ 1p_{ 3/2}~ 1p_{ 3/2}~ 1p_{ 1/2}~ 1p_{ 1/2}$ &  0 &  1 & -0.849 \\
 $ 1p_{ 1/2}~ 1p_{ 1/2}~ 1p_{ 1/2}~ 1p_{ 1/2}$ &  0 &  1 & -0.019 \\
 $ 0f_{ 7/2}~ 0f_{ 5/2}~ 0f_{ 7/2}~ 0f_{ 5/2}$ &  1 &  1 &  0.588 \\
 $ 0f_{ 7/2}~ 0f_{ 5/2}~ 0f_{ 5/2}~ 1p_{ 3/2}$ &  1 &  1 & -0.213 \\
 $ 0f_{ 7/2}~ 0f_{ 5/2}~ 1p_{ 3/2}~ 1p_{ 1/2}$ &  1 &  1 & -0.057 \\
 $ 0f_{ 5/2}~ 1p_{ 3/2}~ 0f_{ 5/2}~ 1p_{ 3/2}$ &  1 &  1 &  0.494 \\
 $ 0f_{ 5/2}~ 1p_{ 3/2}~ 1p_{ 3/2}~ 1p_{ 1/2}$ &  1 &  1 & -0.107 \\
 $ 1p_{ 3/2}~ 1p_{ 1/2}~ 1p_{ 3/2}~ 1p_{ 1/2}$ &  1 &  1 &  0.660 \\
 $ 0f_{ 7/2}~ 0f_{ 7/2}~ 0f_{ 7/2}~ 0f_{ 7/2}$ &  2 &  1 & -0.573 \\
 $ 0f_{ 7/2}~ 0f_{ 7/2}~ 0f_{ 7/2}~ 0f_{ 5/2}$ &  2 &  1 &  0.175 \\
 $ 0f_{ 7/2}~ 0f_{ 7/2}~ 0f_{ 7/2}~ 1p_{ 3/2}$ &  2 &  1 & -0.677 \\
 $ 0f_{ 7/2}~ 0f_{ 7/2}~ 0f_{ 5/2}~ 0f_{ 5/2}$ &  2 &  1 & -0.705 \\
 $ 0f_{ 7/2}~ 0f_{ 7/2}~ 0f_{ 5/2}~ 1p_{ 3/2}$ &  2 &  1 &  0.565 \\
 $ 0f_{ 7/2}~ 0f_{ 7/2}~ 0f_{ 5/2}~ 1p_{ 1/2}$ &  2 &  1 & -0.740 \\
 $ 0f_{ 7/2}~ 0f_{ 7/2}~ 1p_{ 3/2}~ 1p_{ 3/2}$ &  2 &  1 & -0.399 \\
 $ 0f_{ 7/2}~ 0f_{ 7/2}~ 1p_{ 3/2}~ 1p_{ 1/2}$ &  2 &  1 & -0.468 \\
 $ 0f_{ 7/2}~ 0f_{ 5/2}~ 0f_{ 7/2}~ 0f_{ 5/2}$ &  2 &  1 &  0.215 \\
 $ 0f_{ 7/2}~ 0f_{ 5/2}~ 0f_{ 7/2}~ 1p_{ 3/2}$ &  2 &  1 & -0.040 \\
 $ 0f_{ 7/2}~ 0f_{ 5/2}~ 0f_{ 5/2}~ 0f_{ 5/2}$ &  2 &  1 & -0.631 \\
 $ 0f_{ 7/2}~ 0f_{ 5/2}~ 0f_{ 5/2}~ 1p_{ 3/2}$ &  2 &  1 &  0.358 \\
 $ 0f_{ 7/2}~ 0f_{ 5/2}~ 0f_{ 5/2}~ 1p_{ 1/2}$ &  2 &  1 & -0.409 \\
 $ 0f_{ 7/2}~ 0f_{ 5/2}~ 1p_{ 3/2}~ 1p_{ 3/2}$ &  2 &  1 & -0.065 \\
 $ 0f_{ 7/2}~ 0f_{ 5/2}~ 1p_{ 3/2}~ 1p_{ 1/2}$ &  2 &  1 & -0.149 \\
 $ 0f_{ 7/2}~ 1p_{ 3/2}~ 0f_{ 7/2}~ 1p_{ 3/2}$ &  2 &  1 & -0.652 \\
 $ 0f_{ 7/2}~ 1p_{ 3/2}~ 0f_{ 5/2}~ 0f_{ 5/2}$ &  2 &  1 & -0.644 \\
 $ 0f_{ 7/2}~ 1p_{ 3/2}~ 0f_{ 5/2}~ 1p_{ 3/2}$ &  2 &  1 &  0.400 \\
 $ 0f_{ 7/2}~ 1p_{ 3/2}~ 0f_{ 5/2}~ 1p_{ 1/2}$ &  2 &  1 & -1.146 \\
 $ 0f_{ 7/2}~ 1p_{ 3/2}~ 1p_{ 3/2}~ 1p_{ 3/2}$ &  2 &  1 & -0.553 \\
 $ 0f_{ 7/2}~ 1p_{ 3/2}~ 1p_{ 3/2}~ 1p_{ 1/2}$ &  2 &  1 & -0.833 \\
 $ 0f_{ 5/2}~ 0f_{ 5/2}~ 0f_{ 5/2}~ 0f_{ 5/2}$ &  2 &  1 & -0.034 \\
 $ 0f_{ 5/2}~ 0f_{ 5/2}~ 0f_{ 5/2}~ 1p_{ 3/2}$ &  2 &  1 &  0.084 \\
 $ 0f_{ 5/2}~ 0f_{ 5/2}~ 0f_{ 5/2}~ 1p_{ 1/2}$ &  2 &  1 & -0.268 \\
 $ 0f_{ 5/2}~ 0f_{ 5/2}~ 1p_{ 3/2}~ 1p_{ 3/2}$ &  2 &  1 & -0.214 \\
 $ 0f_{ 5/2}~ 0f_{ 5/2}~ 1p_{ 3/2}~ 1p_{ 1/2}$ &  2 &  1 & -0.407 \\
 $ 0f_{ 5/2}~ 1p_{ 3/2}~ 0f_{ 5/2}~ 1p_{ 3/2}$ &  2 &  1 &  0.444 \\
 $ 0f_{ 5/2}~ 1p_{ 3/2}~ 0f_{ 5/2}~ 1p_{ 1/2}$ &  2 &  1 &  0.565 \\
 $ 0f_{ 5/2}~ 1p_{ 3/2}~ 1p_{ 3/2}~ 1p_{ 3/2}$ &  2 &  1 &  0.251 \\
 $ 0f_{ 5/2}~ 1p_{ 3/2}~ 1p_{ 3/2}~ 1p_{ 1/2}$ &  2 &  1 &  0.292 \\
 $ 0f_{ 5/2}~ 1p_{ 1/2}~ 0f_{ 5/2}~ 1p_{ 1/2}$ &  2 &  1 &  0.038 \\
\end{tabular}
\end{ruledtabular}
\label{tablepp}
\end{table}

\begin{table}[H]
\begin{ruledtabular}
\begin{tabular}{cccc}
 $ 0f_{ 5/2}~ 1p_{ 1/2}~ 1p_{ 3/2}~ 1p_{ 3/2}$ &  2 &  1 & -0.417 \\
 $ 0f_{ 5/2}~ 1p_{ 1/2}~ 1p_{ 3/2}~ 1p_{ 1/2}$ &  2 &  1 & -0.508 \\
 $ 1p_{ 3/2}~ 1p_{ 3/2}~ 1p_{ 3/2}~ 1p_{ 3/2}$ &  2 &  1 &  0.221 \\
 $ 1p_{ 3/2}~ 1p_{ 3/2}~ 1p_{ 3/2}~ 1p_{ 1/2}$ &  2 &  1 & -0.536 \\
 $ 1p_{ 3/2}~ 1p_{ 1/2}~ 1p_{ 3/2}~ 1p_{ 1/2}$ &  2 &  1 &  0.006 \\
 $ 0f_{ 7/2}~ 0f_{ 5/2}~ 0f_{ 7/2}~ 0f_{ 5/2}$ &  3 &  1 &  0.739 \\
 $ 0f_{ 7/2}~ 0f_{ 5/2}~ 0f_{ 7/2}~ 1p_{ 3/2}$ &  3 &  1 & -0.259 \\
 $ 0f_{ 7/2}~ 0f_{ 5/2}~ 0f_{ 7/2}~ 1p_{ 1/2}$ &  3 &  1 &  0.168 \\
 $ 0f_{ 7/2}~ 0f_{ 5/2}~ 0f_{ 5/2}~ 1p_{ 3/2}$ &  3 &  1 & -0.177 \\
 $ 0f_{ 7/2}~ 0f_{ 5/2}~ 0f_{ 5/2}~ 1p_{ 1/2}$ &  3 &  1 & -0.062 \\
 $ 0f_{ 7/2}~ 1p_{ 3/2}~ 0f_{ 7/2}~ 1p_{ 3/2}$ &  3 &  1 &  0.582 \\
 $ 0f_{ 7/2}~ 1p_{ 3/2}~ 0f_{ 7/2}~ 1p_{ 1/2}$ &  3 &  1 & -0.033 \\
 $ 0f_{ 7/2}~ 1p_{ 3/2}~ 0f_{ 5/2}~ 1p_{ 3/2}$ &  3 &  1 &  0.079 \\
 $ 0f_{ 7/2}~ 1p_{ 3/2}~ 0f_{ 5/2}~ 1p_{ 1/2}$ &  3 &  1 & -0.073 \\
 $ 0f_{ 7/2}~ 1p_{ 1/2}~ 0f_{ 7/2}~ 1p_{ 1/2}$ &  3 &  1 &  0.670 \\
 $ 0f_{ 7/2}~ 1p_{ 1/2}~ 0f_{ 5/2}~ 1p_{ 3/2}$ &  3 &  1 &  0.155 \\
 $ 0f_{ 7/2}~ 1p_{ 1/2}~ 0f_{ 5/2}~ 1p_{ 1/2}$ &  3 &  1 & -0.076 \\
 $ 0f_{ 5/2}~ 1p_{ 3/2}~ 0f_{ 5/2}~ 1p_{ 3/2}$ &  3 &  1 &  0.578 \\
 $ 0f_{ 5/2}~ 1p_{ 3/2}~ 0f_{ 5/2}~ 1p_{ 1/2}$ &  3 &  1 &  0.050 \\
 $ 0f_{ 5/2}~ 1p_{ 1/2}~ 0f_{ 5/2}~ 1p_{ 1/2}$ &  3 &  1 &  0.608 \\
 $ 0f_{ 7/2}~ 0f_{ 7/2}~ 0f_{ 7/2}~ 0f_{ 7/2}$ &  4 &  1 &  0.248 \\
 $ 0f_{ 7/2}~ 0f_{ 7/2}~ 0f_{ 7/2}~ 0f_{ 5/2}$ &  4 &  1 & -0.571 \\
 $ 0f_{ 7/2}~ 0f_{ 7/2}~ 0f_{ 7/2}~ 1p_{ 3/2}$ &  4 &  1 & -0.453 \\
 $ 0f_{ 7/2}~ 0f_{ 7/2}~ 0f_{ 7/2}~ 1p_{ 1/2}$ &  4 &  1 & -0.460 \\
 $ 0f_{ 7/2}~ 0f_{ 7/2}~ 0f_{ 5/2}~ 0f_{ 5/2}$ &  4 &  1 & -0.536 \\
 $ 0f_{ 7/2}~ 0f_{ 7/2}~ 0f_{ 5/2}~ 1p_{ 3/2}$ &  4 &  1 &  0.710 \\
 $ 0f_{ 7/2}~ 0f_{ 5/2}~ 0f_{ 7/2}~ 0f_{ 5/2}$ &  4 &  1 &  0.534 \\
 $ 0f_{ 7/2}~ 0f_{ 5/2}~ 0f_{ 7/2}~ 1p_{ 3/2}$ &  4 &  1 & -0.255 \\
 $ 0f_{ 7/2}~ 0f_{ 5/2}~ 0f_{ 7/2}~ 1p_{ 1/2}$ &  4 &  1 & -0.085 \\
 $ 0f_{ 7/2}~ 0f_{ 5/2}~ 0f_{ 5/2}~ 0f_{ 5/2}$ &  4 &  1 & -0.482 \\
 $ 0f_{ 7/2}~ 0f_{ 5/2}~ 0f_{ 5/2}~ 1p_{ 3/2}$ &  4 &  1 &  0.788 \\
 $ 0f_{ 7/2}~ 1p_{ 3/2}~ 0f_{ 7/2}~ 1p_{ 3/2}$ &  4 &  1 &  0.276 \\
 $ 0f_{ 7/2}~ 1p_{ 3/2}~ 0f_{ 7/2}~ 1p_{ 1/2}$ &  4 &  1 & -0.760 \\
 $ 0f_{ 7/2}~ 1p_{ 3/2}~ 0f_{ 5/2}~ 0f_{ 5/2}$ &  4 &  1 & -0.270 \\
 $ 0f_{ 7/2}~ 1p_{ 3/2}~ 0f_{ 5/2}~ 1p_{ 3/2}$ &  4 &  1 &  0.801 \\
 $ 0f_{ 7/2}~ 1p_{ 1/2}~ 0f_{ 7/2}~ 1p_{ 1/2}$ &  4 &  1 &  0.125 \\
 $ 0f_{ 7/2}~ 1p_{ 1/2}~ 0f_{ 5/2}~ 0f_{ 5/2}$ &  4 &  1 & -0.340 \\
 $ 0f_{ 7/2}~ 1p_{ 1/2}~ 0f_{ 5/2}~ 1p_{ 3/2}$ &  4 &  1 &  0.825 \\
 $ 0f_{ 5/2}~ 0f_{ 5/2}~ 0f_{ 5/2}~ 0f_{ 5/2}$ &  4 &  1 &  0.335 \\
 $ 0f_{ 5/2}~ 0f_{ 5/2}~ 0f_{ 5/2}~ 1p_{ 3/2}$ &  4 &  1 &  0.247 \\
 $ 0f_{ 5/2}~ 1p_{ 3/2}~ 0f_{ 5/2}~ 1p_{ 3/2}$ &  4 &  1 & -0.106 \\
 $ 0f_{ 7/2}~ 0f_{ 5/2}~ 0f_{ 7/2}~ 0f_{ 5/2}$ &  5 &  1 &  0.898 \\
 $ 0f_{ 7/2}~ 0f_{ 5/2}~ 0f_{ 7/2}~ 1p_{ 3/2}$ &  5 &  1 & -0.233 \\
 $ 0f_{ 7/2}~ 1p_{ 3/2}~ 0f_{ 7/2}~ 1p_{ 3/2}$ &  5 &  1 &  1.085 \\
 $ 0f_{ 7/2}~ 0f_{ 7/2}~ 0f_{ 7/2}~ 0f_{ 7/2}$ &  6 &  1 &  0.673 \\
 $ 0f_{ 7/2}~ 0f_{ 7/2}~ 0f_{ 7/2}~ 0f_{ 5/2}$ &  6 &  1 & -1.207 \\
 $ 0f_{ 7/2}~ 0f_{ 5/2}~ 0f_{ 7/2}~ 0f_{ 5/2}$ &  6 &  1 & -0.716 \\
\end{tabular}
\end{ruledtabular}
\end{table}

\begin{table}[H]
\caption{Same as in Table \ref{tablenn}, but for proton-neutron TBME.}
\begin{ruledtabular}
\begin{tabular}{cccc}
$n_a l_a j_a ~ n_b l_b j_b ~ n_c l_c j_c ~ n_d l_d j_d $ & $J$ & $T_z$
  &  TBME \\
\colrule
 $ 0f_{ 7/2}~ 0f_{ 7/2}~ 0f_{ 7/2}~ 0f_{ 7/2}$ &  0 &  0 & -1.943 \\
 $ 0f_{ 7/2}~ 0f_{ 7/2}~ 0f_{ 5/2}~ 0f_{ 5/2}$ &  0 &  0 & -2.690 \\
 $ 0f_{ 7/2}~ 0f_{ 7/2}~ 1p_{ 3/2}~ 1p_{ 3/2}$ &  0 &  0 & -1.168 \\
 $ 0f_{ 7/2}~ 0f_{ 7/2}~ 1p_{ 1/2}~ 1p_{ 1/2}$ &  0 &  0 & -0.954 \\
 $ 0f_{ 5/2}~ 0f_{ 5/2}~ 0f_{ 5/2}~ 0f_{ 5/2}$ &  0 &  0 & -0.902 \\
 $ 0f_{ 5/2}~ 0f_{ 5/2}~ 1p_{ 3/2}~ 1p_{ 3/2}$ &  0 &  0 & -0.887 \\
 $ 0f_{ 5/2}~ 0f_{ 5/2}~ 1p_{ 1/2}~ 1p_{ 1/2}$ &  0 &  0 & -0.400 \\
 $ 1p_{ 3/2}~ 1p_{ 3/2}~ 1p_{ 3/2}~ 1p_{ 3/2}$ &  0 &  0 & -1.061 \\
 $ 1p_{ 3/2}~ 1p_{ 3/2}~ 1p_{ 1/2}~ 1p_{ 1/2}$ &  0 &  0 & -1.181 \\
\end{tabular}
\end{ruledtabular}
\label{tablepn}
\end{table}

\begin{table}[H]
\begin{ruledtabular}
\begin{tabular}{cccc}
 $ 1p_{ 1/2}~ 1p_{ 1/2}~ 1p_{ 1/2}~ 1p_{ 1/2}$ &  0 &  0 & -0.182 \\
 $ 0f_{ 7/2}~ 0f_{ 7/2}~ 0f_{ 7/2}~ 0f_{ 7/2}$ &  1 &  0 & -1.236 \\
 $ 0f_{ 7/2}~ 0f_{ 7/2}~ 0f_{ 7/2}~ 0f_{ 5/2}$ &  1 &  0 &  1.949 \\
 $ 0f_{ 7/2}~ 0f_{ 7/2}~ 0f_{ 5/2}~ 0f_{ 7/2}$ &  1 &  0 & -1.893 \\
 $ 0f_{ 7/2}~ 0f_{ 7/2}~ 0f_{ 5/2}~ 0f_{ 5/2}$ &  1 &  0 &  1.835 \\
 $ 0f_{ 7/2}~ 0f_{ 7/2}~ 0f_{ 5/2}~ 1p_{ 3/2}$ &  1 &  0 &  0.480 \\
 $ 0f_{ 7/2}~ 0f_{ 7/2}~ 1p_{ 3/2}~ 0f_{ 5/2}$ &  1 &  0 & -0.429 \\
 $ 0f_{ 7/2}~ 0f_{ 7/2}~ 1p_{ 3/2}~ 1p_{ 3/2}$ &  1 &  0 & -0.748 \\
 $ 0f_{ 7/2}~ 0f_{ 7/2}~ 1p_{ 3/2}~ 1p_{ 1/2}$ &  1 &  0 &  0.583 \\
 $ 0f_{ 7/2}~ 0f_{ 7/2}~ 1p_{ 1/2}~ 1p_{ 3/2}$ &  1 &  0 & -0.586 \\
 $ 0f_{ 7/2}~ 0f_{ 7/2}~ 1p_{ 1/2}~ 1p_{ 1/2}$ &  1 &  0 &  0.395 \\
 $ 0f_{ 7/2}~ 0f_{ 5/2}~ 0f_{ 7/2}~ 0f_{ 5/2}$ &  1 &  0 & -2.369 \\
 $ 0f_{ 7/2}~ 0f_{ 5/2}~ 0f_{ 5/2}~ 0f_{ 7/2}$ &  1 &  0 &  2.520 \\
 $ 0f_{ 7/2}~ 0f_{ 5/2}~ 0f_{ 5/2}~ 0f_{ 5/2}$ &  1 &  0 & -0.479 \\
 $ 0f_{ 7/2}~ 0f_{ 5/2}~ 0f_{ 5/2}~ 1p_{ 3/2}$ &  1 &  0 &  0.604 \\
 $ 0f_{ 7/2}~ 0f_{ 5/2}~ 1p_{ 3/2}~ 0f_{ 5/2}$ &  1 &  0 & -0.697 \\
 $ 0f_{ 7/2}~ 0f_{ 5/2}~ 1p_{ 3/2}~ 1p_{ 3/2}$ &  1 &  0 &  1.066 \\
 $ 0f_{ 7/2}~ 0f_{ 5/2}~ 1p_{ 3/2}~ 1p_{ 1/2}$ &  1 &  0 & -1.126 \\
 $ 0f_{ 7/2}~ 0f_{ 5/2}~ 1p_{ 1/2}~ 1p_{ 3/2}$ &  1 &  0 &  1.085 \\
 $ 0f_{ 7/2}~ 0f_{ 5/2}~ 1p_{ 1/2}~ 1p_{ 1/2}$ &  1 &  0 &  0.105 \\
 $ 0f_{ 5/2}~ 0f_{ 7/2}~ 0f_{ 5/2}~ 0f_{ 7/2}$ &  1 &  0 & -2.204 \\
 $ 0f_{ 5/2}~ 0f_{ 7/2}~ 0f_{ 5/2}~ 0f_{ 5/2}$ &  1 &  0 &  0.562 \\
 $ 0f_{ 5/2}~ 0f_{ 7/2}~ 0f_{ 5/2}~ 1p_{ 3/2}$ &  1 &  0 & -0.666 \\
 $ 0f_{ 5/2}~ 0f_{ 7/2}~ 1p_{ 3/2}~ 0f_{ 5/2}$ &  1 &  0 &  0.602 \\
 $ 0f_{ 5/2}~ 0f_{ 7/2}~ 1p_{ 3/2}~ 1p_{ 3/2}$ &  1 &  0 & -1.061 \\
 $ 0f_{ 5/2}~ 0f_{ 7/2}~ 1p_{ 3/2}~ 1p_{ 1/2}$ &  1 &  0 &  1.031 \\
 $ 0f_{ 5/2}~ 0f_{ 7/2}~ 1p_{ 1/2}~ 1p_{ 3/2}$ &  1 &  0 & -1.041 \\
 $ 0f_{ 5/2}~ 0f_{ 7/2}~ 1p_{ 1/2}~ 1p_{ 1/2}$ &  1 &  0 & -0.089 \\
 $ 0f_{ 5/2}~ 0f_{ 5/2}~ 0f_{ 5/2}~ 0f_{ 5/2}$ &  1 &  0 & -0.654 \\
 $ 0f_{ 5/2}~ 0f_{ 5/2}~ 0f_{ 5/2}~ 1p_{ 3/2}$ &  1 &  0 & -0.562 \\
 $ 0f_{ 5/2}~ 0f_{ 5/2}~ 1p_{ 3/2}~ 0f_{ 5/2}$ &  1 &  0 &  0.524 \\
 $ 0f_{ 5/2}~ 0f_{ 5/2}~ 1p_{ 3/2}~ 1p_{ 3/2}$ &  1 &  0 &  0.229 \\
 $ 0f_{ 5/2}~ 0f_{ 5/2}~ 1p_{ 3/2}~ 1p_{ 1/2}$ &  1 &  0 & -0.034 \\
 $ 0f_{ 5/2}~ 0f_{ 5/2}~ 1p_{ 1/2}~ 1p_{ 3/2}$ &  1 &  0 &  0.035 \\
 $ 0f_{ 5/2}~ 0f_{ 5/2}~ 1p_{ 1/2}~ 1p_{ 1/2}$ &  1 &  0 & -0.278 \\
 $ 0f_{ 5/2}~ 1p_{ 3/2}~ 0f_{ 5/2}~ 1p_{ 3/2}$ &  1 &  0 & -1.597 \\
 $ 0f_{ 5/2}~ 1p_{ 3/2}~ 1p_{ 3/2}~ 0f_{ 5/2}$ &  1 &  0 &  1.734 \\
 $ 0f_{ 5/2}~ 1p_{ 3/2}~ 1p_{ 3/2}~ 1p_{ 3/2}$ &  1 &  0 &  0.204 \\
 $ 0f_{ 5/2}~ 1p_{ 3/2}~ 1p_{ 3/2}~ 1p_{ 1/2}$ &  1 &  0 &  0.545 \\
 $ 0f_{ 5/2}~ 1p_{ 3/2}~ 1p_{ 1/2}~ 1p_{ 3/2}$ &  1 &  0 & -0.576 \\
 $ 0f_{ 5/2}~ 1p_{ 3/2}~ 1p_{ 1/2}~ 1p_{ 1/2}$ &  1 &  0 & -0.995 \\
 $ 1p_{ 3/2}~ 0f_{ 5/2}~ 1p_{ 3/2}~ 0f_{ 5/2}$ &  1 &  0 & -1.483 \\
 $ 1p_{ 3/2}~ 0f_{ 5/2}~ 1p_{ 3/2}~ 1p_{ 3/2}$ &  1 &  0 & -0.213 \\
 $ 1p_{ 3/2}~ 0f_{ 5/2}~ 1p_{ 3/2}~ 1p_{ 1/2}$ &  1 &  0 & -0.538 \\
 $ 1p_{ 3/2}~ 0f_{ 5/2}~ 1p_{ 1/2}~ 1p_{ 3/2}$ &  1 &  0 &  0.484 \\
 $ 1p_{ 3/2}~ 0f_{ 5/2}~ 1p_{ 1/2}~ 1p_{ 1/2}$ &  1 &  0 &  0.965 \\
 $ 1p_{ 3/2}~ 1p_{ 3/2}~ 1p_{ 3/2}~ 1p_{ 3/2}$ &  1 &  0 & -0.442 \\
 $ 1p_{ 3/2}~ 1p_{ 3/2}~ 1p_{ 3/2}~ 1p_{ 1/2}$ &  1 &  0 &  1.390 \\
 $ 1p_{ 3/2}~ 1p_{ 3/2}~ 1p_{ 1/2}~ 1p_{ 3/2}$ &  1 &  0 & -1.359 \\
 $ 1p_{ 3/2}~ 1p_{ 3/2}~ 1p_{ 1/2}~ 1p_{ 1/2}$ &  1 &  0 &  1.037 \\
 $ 1p_{ 3/2}~ 1p_{ 1/2}~ 1p_{ 3/2}~ 1p_{ 1/2}$ &  1 &  0 & -0.840 \\
 $ 1p_{ 3/2}~ 1p_{ 1/2}~ 1p_{ 1/2}~ 1p_{ 3/2}$ &  1 &  0 &  1.171 \\
 $ 1p_{ 3/2}~ 1p_{ 1/2}~ 1p_{ 1/2}~ 1p_{ 1/2}$ &  1 &  0 &  0.412 \\
 $ 1p_{ 1/2}~ 1p_{ 3/2}~ 1p_{ 1/2}~ 1p_{ 3/2}$ &  1 &  0 & -0.741 \\
 $ 1p_{ 1/2}~ 1p_{ 3/2}~ 1p_{ 1/2}~ 1p_{ 1/2}$ &  1 &  0 & -0.405 \\
 $ 1p_{ 1/2}~ 1p_{ 1/2}~ 1p_{ 1/2}~ 1p_{ 1/2}$ &  1 &  0 & -0.684 \\
 $ 0f_{ 7/2}~ 0f_{ 7/2}~ 0f_{ 7/2}~ 0f_{ 7/2}$ &  2 &  0 & -0.988 \\
 $ 0f_{ 7/2}~ 0f_{ 7/2}~ 0f_{ 7/2}~ 0f_{ 5/2}$ &  2 &  0 &  0.110 \\
 $ 0f_{ 7/2}~ 0f_{ 7/2}~ 0f_{ 7/2}~ 1p_{ 3/2}$ &  2 &  0 & -0.478 \\
 $ 0f_{ 7/2}~ 0f_{ 7/2}~ 0f_{ 5/2}~ 0f_{ 7/2}$ &  2 &  0 & -0.118 \\
 $ 0f_{ 7/2}~ 0f_{ 7/2}~ 0f_{ 5/2}~ 0f_{ 5/2}$ &  2 &  0 & -0.715 \\
 $ 0f_{ 7/2}~ 0f_{ 7/2}~ 0f_{ 5/2}~ 1p_{ 3/2}$ &  2 &  0 &  0.354 \\
 $ 0f_{ 7/2}~ 0f_{ 7/2}~ 0f_{ 5/2}~ 1p_{ 1/2}$ &  2 &  0 & -0.454 \\
\end{tabular}
\end{ruledtabular}
\end{table}

\begin{table}[H]
\begin{ruledtabular}
\begin{tabular}{cccc}
 $ 0f_{ 7/2}~ 0f_{ 7/2}~ 1p_{ 3/2}~ 0f_{ 7/2}$ &  2 &  0 & -0.452 \\
 $ 0f_{ 7/2}~ 0f_{ 7/2}~ 1p_{ 3/2}~ 0f_{ 5/2}$ &  2 &  0 & -0.334 \\
 $ 0f_{ 7/2}~ 0f_{ 7/2}~ 1p_{ 3/2}~ 1p_{ 3/2}$ &  2 &  0 & -0.370 \\
 $ 0f_{ 7/2}~ 0f_{ 7/2}~ 1p_{ 3/2}~ 1p_{ 1/2}$ &  2 &  0 & -0.293 \\
 $ 0f_{ 7/2}~ 0f_{ 7/2}~ 1p_{ 1/2}~ 0f_{ 5/2}$ &  2 &  0 & -0.414 \\
 $ 0f_{ 7/2}~ 0f_{ 7/2}~ 1p_{ 1/2}~ 1p_{ 3/2}$ &  2 &  0 &  0.289 \\
 $ 0f_{ 7/2}~ 0f_{ 5/2}~ 0f_{ 7/2}~ 0f_{ 5/2}$ &  2 &  0 & -1.807 \\
 $ 0f_{ 7/2}~ 0f_{ 5/2}~ 0f_{ 7/2}~ 1p_{ 3/2}$ &  2 &  0 & -0.679 \\
 $ 0f_{ 7/2}~ 0f_{ 5/2}~ 0f_{ 5/2}~ 0f_{ 7/2}$ &  2 &  0 & -1.753 \\
 $ 0f_{ 7/2}~ 0f_{ 5/2}~ 0f_{ 5/2}~ 0f_{ 5/2}$ &  2 &  0 & -0.471 \\
 $ 0f_{ 7/2}~ 0f_{ 5/2}~ 0f_{ 5/2}~ 1p_{ 3/2}$ &  2 &  0 & -0.564 \\
 $ 0f_{ 7/2}~ 0f_{ 5/2}~ 0f_{ 5/2}~ 1p_{ 1/2}$ &  2 &  0 &  0.287 \\
 $ 0f_{ 7/2}~ 0f_{ 5/2}~ 1p_{ 3/2}~ 0f_{ 7/2}$ &  2 &  0 &  0.683 \\
 $ 0f_{ 7/2}~ 0f_{ 5/2}~ 1p_{ 3/2}~ 0f_{ 5/2}$ &  2 &  0 & -0.799 \\
 $ 0f_{ 7/2}~ 0f_{ 5/2}~ 1p_{ 3/2}~ 1p_{ 3/2}$ &  2 &  0 & -0.030 \\
 $ 0f_{ 7/2}~ 0f_{ 5/2}~ 1p_{ 3/2}~ 1p_{ 1/2}$ &  2 &  0 & -0.663 \\
 $ 0f_{ 7/2}~ 0f_{ 5/2}~ 1p_{ 1/2}~ 0f_{ 5/2}$ &  2 &  0 & -0.685 \\
 $ 0f_{ 7/2}~ 0f_{ 5/2}~ 1p_{ 1/2}~ 1p_{ 3/2}$ &  2 &  0 & -0.521 \\
 $ 1p_{ 3/2}~ 1p_{ 3/2}~ 1p_{ 1/2}~ 1p_{ 3/2}$ &  2 &  0 &  0.480 \\
 $ 0f_{ 7/2}~ 1p_{ 3/2}~ 0f_{ 7/2}~ 1p_{ 3/2}$ &  2 &  0 & -0.777 \\
 $ 0f_{ 7/2}~ 1p_{ 3/2}~ 0f_{ 5/2}~ 0f_{ 7/2}$ &  2 &  0 & -0.794 \\
 $ 0f_{ 7/2}~ 1p_{ 3/2}~ 0f_{ 5/2}~ 0f_{ 5/2}$ &  2 &  0 & -0.500 \\
 $ 0f_{ 7/2}~ 1p_{ 3/2}~ 0f_{ 5/2}~ 1p_{ 3/2}$ &  2 &  0 & -0.630 \\
 $ 0f_{ 7/2}~ 1p_{ 3/2}~ 0f_{ 5/2}~ 1p_{ 1/2}$ &  2 &  0 &  0.152 \\
 $ 0f_{ 7/2}~ 1p_{ 3/2}~ 1p_{ 3/2}~ 0f_{ 7/2}$ &  2 &  0 & -0.252 \\
 $ 0f_{ 7/2}~ 1p_{ 3/2}~ 1p_{ 3/2}~ 0f_{ 5/2}$ &  2 &  0 & -1.066 \\
 $ 0f_{ 7/2}~ 1p_{ 3/2}~ 1p_{ 3/2}~ 1p_{ 3/2}$ &  2 &  0 & -0.390 \\
 $ 0f_{ 7/2}~ 1p_{ 3/2}~ 1p_{ 3/2}~ 1p_{ 1/2}$ &  2 &  0 & -0.831 \\
 $ 0f_{ 7/2}~ 1p_{ 3/2}~ 1p_{ 1/2}~ 0f_{ 5/2}$ &  2 &  0 & -1.385 \\
 $ 0f_{ 7/2}~ 1p_{ 3/2}~ 1p_{ 1/2}~ 1p_{ 3/2}$ &  2 &  0 & -0.153 \\
 $ 0f_{ 5/2}~ 0f_{ 7/2}~ 0f_{ 5/2}~ 0f_{ 7/2}$ &  2 &  0 & -1.701 \\
 $ 0f_{ 5/2}~ 0f_{ 7/2}~ 0f_{ 5/2}~ 0f_{ 5/2}$ &  2 &  0 &  0.539 \\
 $ 0f_{ 5/2}~ 0f_{ 7/2}~ 0f_{ 5/2}~ 1p_{ 3/2}$ &  2 &  0 & -0.825 \\
 $ 0f_{ 5/2}~ 0f_{ 7/2}~ 0f_{ 5/2}~ 1p_{ 1/2}$ &  2 &  0 &  0.683 \\
 $ 0f_{ 5/2}~ 0f_{ 7/2}~ 1p_{ 3/2}~ 0f_{ 7/2}$ &  2 &  0 &  0.618 \\
 $ 0f_{ 5/2}~ 0f_{ 7/2}~ 1p_{ 3/2}~ 0f_{ 5/2}$ &  2 &  0 & -0.478 \\
 $ 0f_{ 5/2}~ 0f_{ 7/2}~ 1p_{ 3/2}~ 1p_{ 3/2}$ &  2 &  0 &  0.025 \\
 $ 0f_{ 5/2}~ 0f_{ 7/2}~ 1p_{ 3/2}~ 1p_{ 1/2}$ &  2 &  0 & -0.488 \\
 $ 0f_{ 5/2}~ 0f_{ 7/2}~ 1p_{ 1/2}~ 0f_{ 5/2}$ &  2 &  0 & -0.241 \\
 $ 0f_{ 5/2}~ 0f_{ 7/2}~ 1p_{ 1/2}~ 1p_{ 3/2}$ &  2 &  0 & -0.626 \\
 $ 0f_{ 5/2}~ 0f_{ 5/2}~ 0f_{ 5/2}~ 0f_{ 5/2}$ &  2 &  0 & -0.375 \\
 $ 0f_{ 5/2}~ 0f_{ 5/2}~ 0f_{ 5/2}~ 1p_{ 3/2}$ &  2 &  0 &  0.039 \\
 $ 0f_{ 5/2}~ 0f_{ 5/2}~ 0f_{ 5/2}~ 1p_{ 1/2}$ &  2 &  0 & -0.199 \\
 $ 0f_{ 5/2}~ 0f_{ 5/2}~ 1p_{ 3/2}~ 0f_{ 7/2}$ &  2 &  0 & -0.375 \\
 $ 0f_{ 5/2}~ 0f_{ 5/2}~ 1p_{ 3/2}~ 0f_{ 5/2}$ &  2 &  0 & -0.050 \\
 $ 0f_{ 5/2}~ 0f_{ 5/2}~ 1p_{ 3/2}~ 1p_{ 3/2}$ &  2 &  0 & -0.172 \\
 $ 0f_{ 5/2}~ 0f_{ 5/2}~ 1p_{ 3/2}~ 1p_{ 1/2}$ &  2 &  0 & -0.262 \\
 $ 0f_{ 5/2}~ 0f_{ 5/2}~ 1p_{ 1/2}~ 0f_{ 5/2}$ &  2 &  0 & -0.182 \\
 $ 0f_{ 5/2}~ 0f_{ 5/2}~ 1p_{ 1/2}~ 1p_{ 3/2}$ &  2 &  0 &  0.244 \\
 $ 0f_{ 5/2}~ 1p_{ 3/2}~ 0f_{ 5/2}~ 1p_{ 3/2}$ &  2 &  0 & -0.644 \\
 $ 0f_{ 5/2}~ 1p_{ 3/2}~ 0f_{ 5/2}~ 1p_{ 1/2}$ &  2 &  0 &  0.588 \\
 $ 0f_{ 5/2}~ 1p_{ 3/2}~ 1p_{ 3/2}~ 0f_{ 7/2}$ &  2 &  0 &  1.053 \\
 $ 0f_{ 5/2}~ 1p_{ 3/2}~ 1p_{ 3/2}~ 0f_{ 5/2}$ &  2 &  0 & -0.765 \\
 $ 0f_{ 5/2}~ 1p_{ 3/2}~ 1p_{ 3/2}~ 1p_{ 3/2}$ &  2 &  0 &  0.146 \\
 $ 0f_{ 5/2}~ 1p_{ 3/2}~ 1p_{ 3/2}~ 1p_{ 1/2}$ &  2 &  0 & -0.340 \\
 $ 0f_{ 5/2}~ 1p_{ 3/2}~ 1p_{ 1/2}~ 0f_{ 5/2}$ &  2 &  0 &  0.004 \\
 $ 0f_{ 5/2}~ 1p_{ 3/2}~ 1p_{ 1/2}~ 1p_{ 3/2}$ &  2 &  0 & -0.594 \\
 $ 0f_{ 5/2}~ 1p_{ 1/2}~ 0f_{ 5/2}~ 1p_{ 1/2}$ &  2 &  0 & -0.364 \\
 $ 0f_{ 5/2}~ 1p_{ 1/2}~ 1p_{ 3/2}~ 0f_{ 7/2}$ &  2 &  0 & -1.358 \\
 $ 0f_{ 5/2}~ 1p_{ 1/2}~ 1p_{ 3/2}~ 0f_{ 5/2}$ &  2 &  0 & -0.010 \\
 $ 0f_{ 5/2}~ 1p_{ 1/2}~ 1p_{ 3/2}~ 1p_{ 3/2}$ &  2 &  0 & -0.194 \\
 $ 0f_{ 5/2}~ 1p_{ 1/2}~ 1p_{ 3/2}~ 1p_{ 1/2}$ &  2 &  0 &  0.052 \\
 $ 0f_{ 5/2}~ 1p_{ 1/2}~ 1p_{ 1/2}~ 0f_{ 5/2}$ &  2 &  0 &  0.061 \\
\end{tabular}
\end{ruledtabular}
\end{table}

\begin{table}[H]
\begin{ruledtabular}
\begin{tabular}{cccc}
 $ 0f_{ 5/2}~ 1p_{ 1/2}~ 1p_{ 1/2}~ 1p_{ 3/2}$ &  2 &  0 &  0.541 \\
 $ 1p_{ 3/2}~ 0f_{ 7/2}~ 1p_{ 3/2}~ 0f_{ 7/2}$ &  2 &  0 & -0.700 \\
 $ 1p_{ 3/2}~ 0f_{ 7/2}~ 1p_{ 3/2}~ 0f_{ 5/2}$ &  2 &  0 &  0.531 \\
 $ 1p_{ 3/2}~ 0f_{ 7/2}~ 1p_{ 3/2}~ 1p_{ 3/2}$ &  2 &  0 & -0.366 \\
 $ 1p_{ 3/2}~ 0f_{ 7/2}~ 1p_{ 3/2}~ 1p_{ 1/2}$ &  2 &  0 &  0.109 \\
 $ 1p_{ 3/2}~ 0f_{ 7/2}~ 1p_{ 1/2}~ 0f_{ 5/2}$ &  2 &  0 &  0.156 \\
 $ 1p_{ 3/2}~ 0f_{ 7/2}~ 1p_{ 1/2}~ 1p_{ 3/2}$ &  2 &  0 &  0.804 \\
 $ 1p_{ 3/2}~ 0f_{ 5/2}~ 1p_{ 3/2}~ 0f_{ 5/2}$ &  2 &  0 & -0.584 \\
 $ 1p_{ 3/2}~ 0f_{ 5/2}~ 1p_{ 3/2}~ 1p_{ 3/2}$ &  2 &  0 & -0.093 \\
 $ 1p_{ 3/2}~ 0f_{ 5/2}~ 1p_{ 3/2}~ 1p_{ 1/2}$ &  2 &  0 & -0.607 \\
 $ 1p_{ 3/2}~ 0f_{ 5/2}~ 1p_{ 1/2}~ 0f_{ 5/2}$ &  2 &  0 & -0.519 \\
 $ 1p_{ 3/2}~ 0f_{ 5/2}~ 1p_{ 1/2}~ 1p_{ 3/2}$ &  2 &  0 & -0.314 \\
 $ 1p_{ 3/2}~ 1p_{ 3/2}~ 1p_{ 3/2}~ 1p_{ 3/2}$ &  2 &  0 & -0.222 \\
 $ 1p_{ 3/2}~ 1p_{ 3/2}~ 1p_{ 3/2}~ 1p_{ 1/2}$ &  2 &  0 & -0.498 \\
 $ 1p_{ 3/2}~ 1p_{ 3/2}~ 1p_{ 1/2}~ 0f_{ 5/2}$ &  2 &  0 & -0.287 \\
 $ 1p_{ 3/2}~ 1p_{ 1/2}~ 1p_{ 3/2}~ 1p_{ 1/2}$ &  2 &  0 & -1.215 \\
 $ 1p_{ 3/2}~ 1p_{ 1/2}~ 1p_{ 1/2}~ 0f_{ 5/2}$ &  2 &  0 & -0.530 \\
 $ 1p_{ 3/2}~ 1p_{ 1/2}~ 1p_{ 1/2}~ 1p_{ 3/2}$ &  2 &  0 & -0.603 \\
 $ 1p_{ 1/2}~ 0f_{ 5/2}~ 1p_{ 1/2}~ 0f_{ 5/2}$ &  2 &  0 & -0.303 \\
 $ 1p_{ 1/2}~ 0f_{ 5/2}~ 1p_{ 1/2}~ 1p_{ 3/2}$ &  2 &  0 & -0.062 \\
 $ 1p_{ 1/2}~ 1p_{ 3/2}~ 1p_{ 1/2}~ 1p_{ 3/2}$ &  2 &  0 & -1.135 \\
 $ 0f_{ 7/2}~ 0f_{ 7/2}~ 0f_{ 7/2}~ 0f_{ 7/2}$ &  3 &  0 & -0.340 \\
 $ 0f_{ 7/2}~ 0f_{ 7/2}~ 0f_{ 7/2}~ 0f_{ 5/2}$ &  3 &  0 &  1.014 \\
 $ 0f_{ 7/2}~ 0f_{ 7/2}~ 0f_{ 7/2}~ 1p_{ 3/2}$ &  3 &  0 & -0.638 \\
 $ 0f_{ 7/2}~ 0f_{ 7/2}~ 0f_{ 7/2}~ 1p_{ 1/2}$ &  3 &  0 &  0.720 \\
 $ 0f_{ 7/2}~ 0f_{ 7/2}~ 0f_{ 5/2}~ 0f_{ 7/2}$ &  3 &  0 & -0.982 \\
 $ 0f_{ 7/2}~ 0f_{ 7/2}~ 0f_{ 5/2}~ 0f_{ 5/2}$ &  3 &  0 &  0.963 \\
 $ 0f_{ 7/2}~ 0f_{ 7/2}~ 0f_{ 5/2}~ 1p_{ 3/2}$ &  3 &  0 & -0.105 \\
 $ 0f_{ 7/2}~ 0f_{ 7/2}~ 0f_{ 5/2}~ 1p_{ 1/2}$ &  3 &  0 &  0.157 \\
 $ 0f_{ 7/2}~ 0f_{ 7/2}~ 1p_{ 3/2}~ 0f_{ 7/2}$ &  3 &  0 & -0.587 \\
 $ 0f_{ 7/2}~ 0f_{ 7/2}~ 1p_{ 3/2}~ 0f_{ 5/2}$ &  3 &  0 &  0.112 \\
 $ 0f_{ 7/2}~ 0f_{ 7/2}~ 1p_{ 3/2}~ 1p_{ 3/2}$ &  3 &  0 & -0.536 \\
 $ 0f_{ 7/2}~ 0f_{ 7/2}~ 1p_{ 1/2}~ 0f_{ 7/2}$ &  3 &  0 & -0.623 \\
 $ 0f_{ 7/2}~ 0f_{ 7/2}~ 1p_{ 1/2}~ 0f_{ 5/2}$ &  3 &  0 &  0.126 \\
 $ 0f_{ 7/2}~ 0f_{ 5/2}~ 0f_{ 7/2}~ 0f_{ 5/2}$ &  3 &  0 & -0.512 \\
 $ 0f_{ 7/2}~ 0f_{ 5/2}~ 0f_{ 7/2}~ 1p_{ 3/2}$ &  3 &  0 &  0.130 \\
 $ 0f_{ 7/2}~ 0f_{ 5/2}~ 0f_{ 7/2}~ 1p_{ 1/2}$ &  3 &  0 & -0.276 \\
 $ 0f_{ 7/2}~ 0f_{ 5/2}~ 0f_{ 5/2}~ 0f_{ 7/2}$ &  3 &  0 &  0.820 \\
 $ 0f_{ 7/2}~ 0f_{ 5/2}~ 0f_{ 5/2}~ 0f_{ 5/2}$ &  3 &  0 &  0.566 \\
 $ 0f_{ 7/2}~ 0f_{ 5/2}~ 0f_{ 5/2}~ 1p_{ 3/2}$ &  3 &  0 &  0.298 \\
 $ 0f_{ 7/2}~ 0f_{ 5/2}~ 0f_{ 5/2}~ 1p_{ 1/2}$ &  3 &  0 &  0.407 \\
 $ 0f_{ 7/2}~ 0f_{ 5/2}~ 1p_{ 3/2}~ 0f_{ 7/2}$ &  3 &  0 &  0.352 \\
 $ 0f_{ 7/2}~ 0f_{ 5/2}~ 1p_{ 3/2}~ 0f_{ 5/2}$ &  3 &  0 & -0.412 \\
 $ 0f_{ 7/2}~ 0f_{ 5/2}~ 1p_{ 3/2}~ 1p_{ 3/2}$ &  3 &  0 &  0.530 \\
 $ 0f_{ 7/2}~ 0f_{ 5/2}~ 1p_{ 1/2}~ 0f_{ 7/2}$ &  3 &  0 &  0.400 \\
 $ 0f_{ 7/2}~ 0f_{ 5/2}~ 1p_{ 1/2}~ 0f_{ 5/2}$ &  3 &  0 &  0.372 \\
 $ 0f_{ 7/2}~ 1p_{ 3/2}~ 0f_{ 7/2}~ 1p_{ 3/2}$ &  3 &  0 & -0.415 \\
 $ 0f_{ 7/2}~ 1p_{ 3/2}~ 0f_{ 7/2}~ 1p_{ 1/2}$ &  3 &  0 &  0.979 \\
 $ 0f_{ 7/2}~ 1p_{ 3/2}~ 0f_{ 5/2}~ 0f_{ 7/2}$ &  3 &  0 & -0.440 \\
 $ 0f_{ 7/2}~ 1p_{ 3/2}~ 0f_{ 5/2}~ 0f_{ 5/2}$ &  3 &  0 &  0.337 \\
 $ 0f_{ 7/2}~ 1p_{ 3/2}~ 0f_{ 5/2}~ 1p_{ 3/2}$ &  3 &  0 & -0.325 \\
 $ 0f_{ 7/2}~ 1p_{ 3/2}~ 0f_{ 5/2}~ 1p_{ 1/2}$ &  3 &  0 &  0.369 \\
 $ 0f_{ 7/2}~ 1p_{ 3/2}~ 1p_{ 3/2}~ 0f_{ 7/2}$ &  3 &  0 & -0.486 \\
 $ 0f_{ 7/2}~ 1p_{ 3/2}~ 1p_{ 3/2}~ 0f_{ 5/2}$ &  3 &  0 &  0.409 \\
 $ 0f_{ 7/2}~ 1p_{ 3/2}~ 1p_{ 3/2}~ 1p_{ 3/2}$ &  3 &  0 & -0.783 \\
 $ 0f_{ 7/2}~ 1p_{ 3/2}~ 1p_{ 1/2}~ 0f_{ 7/2}$ &  3 &  0 & -0.922 \\
 $ 0f_{ 7/2}~ 1p_{ 3/2}~ 1p_{ 1/2}~ 0f_{ 5/2}$ &  3 &  0 &  0.415 \\
 $ 0f_{ 7/2}~ 1p_{ 1/2}~ 0f_{ 7/2}~ 1p_{ 1/2}$ &  3 &  0 & -0.775 \\
 $ 0f_{ 7/2}~ 1p_{ 1/2}~ 0f_{ 5/2}~ 0f_{ 7/2}$ &  3 &  0 &  0.468 \\
 $ 0f_{ 7/2}~ 1p_{ 1/2}~ 0f_{ 5/2}~ 0f_{ 5/2}$ &  3 &  0 & -0.162 \\
 $ 0f_{ 7/2}~ 1p_{ 1/2}~ 0f_{ 5/2}~ 1p_{ 3/2}$ &  3 &  0 &  0.206 \\
 $ 0f_{ 7/2}~ 1p_{ 1/2}~ 0f_{ 5/2}~ 1p_{ 1/2}$ &  3 &  0 &  0.123 \\
 $ 0f_{ 7/2}~ 1p_{ 1/2}~ 1p_{ 3/2}~ 0f_{ 7/2}$ &  3 &  0 &  0.928 \\
\end{tabular}
\end{ruledtabular}
\end{table}

\begin{table}[H]
\begin{ruledtabular}
\begin{tabular}{cccc}
 $ 0f_{ 7/2}~ 1p_{ 1/2}~ 1p_{ 3/2}~ 0f_{ 5/2}$ &  3 &  0 & -0.022 \\
 $ 0f_{ 7/2}~ 1p_{ 1/2}~ 1p_{ 3/2}~ 1p_{ 3/2}$ &  3 &  0 &  0.871 \\
 $ 0f_{ 7/2}~ 1p_{ 1/2}~ 1p_{ 1/2}~ 0f_{ 7/2}$ &  3 &  0 &  1.013 \\
 $ 0f_{ 7/2}~ 1p_{ 1/2}~ 1p_{ 1/2}~ 0f_{ 5/2}$ &  3 &  0 &  0.186 \\
 $ 0f_{ 5/2}~ 0f_{ 7/2}~ 0f_{ 5/2}~ 0f_{ 7/2}$ &  3 &  0 & -0.440 \\
 $ 0f_{ 5/2}~ 0f_{ 7/2}~ 0f_{ 5/2}~ 0f_{ 5/2}$ &  3 &  0 & -0.551 \\
 $ 0f_{ 5/2}~ 0f_{ 7/2}~ 0f_{ 5/2}~ 1p_{ 3/2}$ &  3 &  0 & -0.419 \\
 $ 0f_{ 5/2}~ 0f_{ 7/2}~ 0f_{ 5/2}~ 1p_{ 1/2}$ &  3 &  0 & -0.419 \\
 $ 0f_{ 5/2}~ 0f_{ 7/2}~ 1p_{ 3/2}~ 0f_{ 7/2}$ &  3 &  0 & -0.119 \\
 $ 0f_{ 5/2}~ 0f_{ 7/2}~ 1p_{ 3/2}~ 0f_{ 5/2}$ &  3 &  0 &  0.260 \\
 $ 0f_{ 5/2}~ 0f_{ 7/2}~ 1p_{ 3/2}~ 1p_{ 3/2}$ &  3 &  0 & -0.528 \\
 $ 0f_{ 5/2}~ 0f_{ 7/2}~ 1p_{ 1/2}~ 0f_{ 7/2}$ &  3 &  0 & -0.200 \\
 $ 0f_{ 5/2}~ 0f_{ 7/2}~ 1p_{ 1/2}~ 0f_{ 5/2}$ &  3 &  0 & -0.376 \\
 $ 0f_{ 5/2}~ 0f_{ 5/2}~ 0f_{ 5/2}~ 0f_{ 5/2}$ &  3 &  0 & -0.704 \\
 $ 0f_{ 5/2}~ 0f_{ 5/2}~ 0f_{ 5/2}~ 1p_{ 3/2}$ &  3 &  0 & -0.336 \\
 $ 0f_{ 5/2}~ 0f_{ 5/2}~ 0f_{ 5/2}~ 1p_{ 1/2}$ &  3 &  0 & -0.586 \\
 $ 0f_{ 5/2}~ 0f_{ 5/2}~ 1p_{ 3/2}~ 0f_{ 7/2}$ &  3 &  0 &  0.335 \\
 $ 0f_{ 5/2}~ 0f_{ 5/2}~ 1p_{ 3/2}~ 0f_{ 5/2}$ &  3 &  0 &  0.303 \\
 $ 0f_{ 5/2}~ 0f_{ 5/2}~ 1p_{ 3/2}~ 1p_{ 3/2}$ &  3 &  0 & -0.084 \\
 $ 0f_{ 5/2}~ 0f_{ 5/2}~ 1p_{ 1/2}~ 0f_{ 7/2}$ &  3 &  0 &  0.199 \\
 $ 0f_{ 5/2}~ 0f_{ 5/2}~ 1p_{ 1/2}~ 0f_{ 5/2}$ &  3 &  0 & -0.529 \\
 $ 0f_{ 5/2}~ 1p_{ 3/2}~ 0f_{ 5/2}~ 1p_{ 3/2}$ &  3 &  0 & -0.326 \\
 $ 0f_{ 5/2}~ 1p_{ 3/2}~ 0f_{ 5/2}~ 1p_{ 1/2}$ &  3 &  0 & -0.706 \\
 $ 0f_{ 5/2}~ 1p_{ 3/2}~ 1p_{ 3/2}~ 0f_{ 7/2}$ &  3 &  0 & -0.359 \\
 $ 0f_{ 5/2}~ 1p_{ 3/2}~ 1p_{ 3/2}~ 0f_{ 5/2}$ &  3 &  0 &  0.542 \\
 $ 0f_{ 5/2}~ 1p_{ 3/2}~ 1p_{ 3/2}~ 1p_{ 3/2}$ &  3 &  0 & -0.331 \\
 $ 0f_{ 5/2}~ 1p_{ 3/2}~ 1p_{ 1/2}~ 0f_{ 7/2}$ &  3 &  0 &  0.062 \\
 $ 0f_{ 5/2}~ 1p_{ 3/2}~ 1p_{ 1/2}~ 0f_{ 5/2}$ &  3 &  0 & -0.773 \\
 $ 0f_{ 5/2}~ 1p_{ 1/2}~ 0f_{ 5/2}~ 1p_{ 1/2}$ &  3 &  0 & -0.915 \\
 $ 0f_{ 5/2}~ 1p_{ 1/2}~ 1p_{ 3/2}~ 0f_{ 7/2}$ &  3 &  0 &  0.454 \\
 $ 0f_{ 5/2}~ 1p_{ 1/2}~ 1p_{ 3/2}~ 0f_{ 5/2}$ &  3 &  0 &  0.788 \\
 $ 0f_{ 5/2}~ 1p_{ 1/2}~ 1p_{ 3/2}~ 1p_{ 3/2}$ &  3 &  0 & -0.053 \\
 $ 0f_{ 5/2}~ 1p_{ 1/2}~ 1p_{ 1/2}~ 0f_{ 7/2}$ &  3 &  0 & -0.105 \\
 $ 0f_{ 5/2}~ 1p_{ 1/2}~ 1p_{ 1/2}~ 0f_{ 5/2}$ &  3 &  0 & -0.970 \\
 $ 1p_{ 3/2}~ 0f_{ 7/2}~ 1p_{ 3/2}~ 0f_{ 7/2}$ &  3 &  0 & -0.400 \\
 $ 1p_{ 3/2}~ 0f_{ 7/2}~ 1p_{ 3/2}~ 0f_{ 5/2}$ &  3 &  0 &  0.290 \\
 $ 1p_{ 3/2}~ 0f_{ 7/2}~ 1p_{ 3/2}~ 1p_{ 3/2}$ &  3 &  0 & -0.728 \\
 $ 1p_{ 3/2}~ 0f_{ 7/2}~ 1p_{ 1/2}~ 0f_{ 7/2}$ &  3 &  0 & -0.861 \\
 $ 1p_{ 3/2}~ 0f_{ 7/2}~ 1p_{ 1/2}~ 0f_{ 5/2}$ &  3 &  0 &  0.333 \\
 $ 1p_{ 3/2}~ 0f_{ 5/2}~ 1p_{ 3/2}~ 0f_{ 5/2}$ &  3 &  0 & -0.320 \\
 $ 1p_{ 3/2}~ 0f_{ 5/2}~ 1p_{ 3/2}~ 1p_{ 3/2}$ &  3 &  0 &  0.256 \\
 $ 1p_{ 3/2}~ 0f_{ 5/2}~ 1p_{ 1/2}~ 0f_{ 7/2}$ &  3 &  0 &  0.117 \\
 $ 1p_{ 3/2}~ 0f_{ 5/2}~ 1p_{ 1/2}~ 0f_{ 5/2}$ &  3 &  0 &  0.650 \\
 $ 1p_{ 3/2}~ 1p_{ 3/2}~ 1p_{ 3/2}~ 1p_{ 3/2}$ &  3 &  0 & -1.649 \\
 $ 1p_{ 3/2}~ 1p_{ 3/2}~ 1p_{ 1/2}~ 0f_{ 7/2}$ &  3 &  0 & -0.922 \\
 $ 1p_{ 3/2}~ 1p_{ 3/2}~ 1p_{ 1/2}~ 0f_{ 5/2}$ &  3 &  0 &  0.037 \\
 $ 1p_{ 1/2}~ 0f_{ 7/2}~ 1p_{ 1/2}~ 0f_{ 7/2}$ &  3 &  0 & -0.676 \\
 $ 1p_{ 1/2}~ 0f_{ 7/2}~ 1p_{ 1/2}~ 0f_{ 5/2}$ &  3 &  0 & -0.134 \\
 $ 1p_{ 1/2}~ 0f_{ 5/2}~ 1p_{ 1/2}~ 0f_{ 5/2}$ &  3 &  0 & -0.799 \\
 $ 0f_{ 7/2}~ 0f_{ 7/2}~ 0f_{ 7/2}~ 0f_{ 7/2}$ &  4 &  0 & -0.178 \\
 $ 0f_{ 7/2}~ 0f_{ 7/2}~ 0f_{ 7/2}~ 0f_{ 5/2}$ &  4 &  0 & -0.414 \\
 $ 0f_{ 7/2}~ 0f_{ 7/2}~ 0f_{ 7/2}~ 1p_{ 3/2}$ &  4 &  0 & -0.318 \\
 $ 0f_{ 7/2}~ 0f_{ 7/2}~ 0f_{ 7/2}~ 1p_{ 1/2}$ &  4 &  0 & -0.331 \\
 $ 0f_{ 7/2}~ 0f_{ 7/2}~ 0f_{ 5/2}~ 0f_{ 7/2}$ &  4 &  0 &  0.399 \\
 $ 0f_{ 7/2}~ 0f_{ 7/2}~ 0f_{ 5/2}~ 0f_{ 5/2}$ &  4 &  0 & -0.525 \\
 $ 0f_{ 7/2}~ 0f_{ 7/2}~ 0f_{ 5/2}~ 1p_{ 3/2}$ &  4 &  0 &  0.435 \\
 $ 0f_{ 7/2}~ 0f_{ 7/2}~ 1p_{ 3/2}~ 0f_{ 7/2}$ &  4 &  0 & -0.291 \\
 $ 0f_{ 7/2}~ 0f_{ 7/2}~ 1p_{ 3/2}~ 0f_{ 5/2}$ &  4 &  0 & -0.404 \\
 $ 0f_{ 7/2}~ 0f_{ 7/2}~ 1p_{ 1/2}~ 0f_{ 7/2}$ &  4 &  0 &  0.278 \\
 $ 0f_{ 7/2}~ 0f_{ 5/2}~ 0f_{ 7/2}~ 0f_{ 5/2}$ &  4 &  0 & -1.219 \\
 $ 0f_{ 7/2}~ 0f_{ 5/2}~ 0f_{ 7/2}~ 1p_{ 3/2}$ &  4 &  0 & -0.234 \\
 $ 0f_{ 7/2}~ 0f_{ 5/2}~ 0f_{ 7/2}~ 1p_{ 1/2}$ &  4 &  0 & -0.572 \\
 $ 0f_{ 7/2}~ 0f_{ 5/2}~ 0f_{ 5/2}~ 0f_{ 7/2}$ &  4 &  0 & -1.244 \\
\end{tabular}
\end{ruledtabular}
\end{table}

\begin{table}[H]
\begin{ruledtabular}
\begin{tabular}{cccc}
 $ 0f_{ 7/2}~ 1p_{ 3/2}~ 1p_{ 3/2}~ 0f_{ 7/2}$ &  4 &  0 &  0.052 \\
 $ 0f_{ 7/2}~ 1p_{ 3/2}~ 1p_{ 3/2}~ 0f_{ 5/2}$ &  4 &  0 & -0.885 \\
 $ 0f_{ 7/2}~ 1p_{ 3/2}~ 1p_{ 1/2}~ 0f_{ 7/2}$ &  4 &  0 &  0.383 \\
 $ 0f_{ 7/2}~ 1p_{ 1/2}~ 0f_{ 7/2}~ 1p_{ 1/2}$ &  4 &  0 & -0.727 \\
 $ 0f_{ 7/2}~ 1p_{ 1/2}~ 0f_{ 5/2}~ 0f_{ 7/2}$ &  4 &  0 & -0.595 \\
 $ 0f_{ 7/2}~ 1p_{ 1/2}~ 0f_{ 5/2}~ 0f_{ 5/2}$ &  4 &  0 & -0.266 \\
 $ 0f_{ 7/2}~ 1p_{ 1/2}~ 0f_{ 5/2}~ 1p_{ 3/2}$ &  4 &  0 & -0.584 \\
 $ 0f_{ 7/2}~ 1p_{ 1/2}~ 1p_{ 3/2}~ 0f_{ 7/2}$ &  4 &  0 & -0.403 \\
 $ 0f_{ 7/2}~ 1p_{ 1/2}~ 1p_{ 3/2}~ 0f_{ 5/2}$ &  4 &  0 & -1.401 \\
 $ 0f_{ 7/2}~ 1p_{ 1/2}~ 1p_{ 1/2}~ 0f_{ 7/2}$ &  4 &  0 & -0.382 \\
 $ 0f_{ 5/2}~ 0f_{ 7/2}~ 0f_{ 5/2}~ 0f_{ 7/2}$ &  4 &  0 & -1.150 \\
 $ 0f_{ 5/2}~ 0f_{ 7/2}~ 0f_{ 5/2}~ 0f_{ 5/2}$ &  4 &  0 &  0.386 \\
 $ 0f_{ 5/2}~ 0f_{ 7/2}~ 0f_{ 5/2}~ 1p_{ 3/2}$ &  4 &  0 & -0.959 \\
 $ 0f_{ 5/2}~ 0f_{ 7/2}~ 1p_{ 3/2}~ 0f_{ 7/2}$ &  4 &  0 &  0.206 \\
 $ 0f_{ 5/2}~ 0f_{ 7/2}~ 1p_{ 3/2}~ 0f_{ 5/2}$ &  4 &  0 & -0.280 \\
 $ 0f_{ 5/2}~ 0f_{ 7/2}~ 1p_{ 1/2}~ 0f_{ 7/2}$ &  4 &  0 & -0.479 \\
 $ 0f_{ 5/2}~ 0f_{ 5/2}~ 0f_{ 5/2}~ 0f_{ 5/2}$ &  4 &  0 & -0.010 \\
 $ 0f_{ 5/2}~ 0f_{ 5/2}~ 0f_{ 5/2}~ 1p_{ 3/2}$ &  4 &  0 &  0.148 \\
 $ 0f_{ 5/2}~ 0f_{ 5/2}~ 1p_{ 3/2}~ 0f_{ 7/2}$ &  4 &  0 & -0.167 \\
 $ 0f_{ 5/2}~ 0f_{ 5/2}~ 1p_{ 3/2}~ 0f_{ 5/2}$ &  4 &  0 & -0.136 \\
 $ 0f_{ 5/2}~ 0f_{ 5/2}~ 1p_{ 1/2}~ 0f_{ 7/2}$ &  4 &  0 &  0.216 \\
 $ 0f_{ 5/2}~ 1p_{ 3/2}~ 0f_{ 5/2}~ 1p_{ 3/2}$ &  4 &  0 & -0.809 \\
 $ 0f_{ 5/2}~ 1p_{ 3/2}~ 1p_{ 3/2}~ 0f_{ 7/2}$ &  4 &  0 &  0.855 \\
 $ 0f_{ 5/2}~ 1p_{ 3/2}~ 1p_{ 3/2}~ 0f_{ 5/2}$ &  4 &  0 & -0.328 \\
 $ 0f_{ 5/2}~ 1p_{ 3/2}~ 1p_{ 1/2}~ 0f_{ 7/2}$ &  4 &  0 & -1.384 \\
 $ 1p_{ 3/2}~ 0f_{ 7/2}~ 1p_{ 3/2}~ 0f_{ 7/2}$ &  4 &  0 & -0.151 \\
 $ 1p_{ 3/2}~ 0f_{ 7/2}~ 1p_{ 3/2}~ 0f_{ 5/2}$ &  4 &  0 &  0.002 \\
 $ 1p_{ 3/2}~ 0f_{ 7/2}~ 1p_{ 1/2}~ 0f_{ 7/2}$ &  4 &  0 &  0.311 \\
 $ 1p_{ 3/2}~ 0f_{ 5/2}~ 1p_{ 3/2}~ 0f_{ 5/2}$ &  4 &  0 & -0.722 \\
 $ 1p_{ 3/2}~ 0f_{ 5/2}~ 1p_{ 1/2}~ 0f_{ 7/2}$ &  4 &  0 & -0.570 \\
 $ 1p_{ 1/2}~ 0f_{ 7/2}~ 1p_{ 1/2}~ 0f_{ 7/2}$ &  4 &  0 & -0.647 \\
 $ 0f_{ 7/2}~ 0f_{ 7/2}~ 0f_{ 7/2}~ 0f_{ 7/2}$ &  5 &  0 & -0.628 \\
 $ 0f_{ 7/2}~ 0f_{ 7/2}~ 0f_{ 7/2}~ 0f_{ 5/2}$ &  5 &  0 &  0.916 \\
 $ 0f_{ 7/2}~ 0f_{ 7/2}~ 0f_{ 7/2}~ 1p_{ 3/2}$ &  5 &  0 & -0.913 \\
 $ 0f_{ 7/2}~ 0f_{ 7/2}~ 0f_{ 5/2}~ 0f_{ 7/2}$ &  5 &  0 & -0.896 \\
 $ 0f_{ 7/2}~ 0f_{ 7/2}~ 0f_{ 5/2}~ 0f_{ 5/2}$ &  5 &  0 &  0.304 \\
 $ 0f_{ 7/2}~ 0f_{ 7/2}~ 1p_{ 3/2}~ 0f_{ 7/2}$ &  5 &  0 & -0.817 \\
 $ 0f_{ 7/2}~ 0f_{ 5/2}~ 0f_{ 7/2}~ 0f_{ 5/2}$ &  5 &  0 &  0.174 \\
 $ 0f_{ 7/2}~ 0f_{ 5/2}~ 0f_{ 7/2}~ 1p_{ 3/2}$ &  5 &  0 &  0.205 \\
 $ 0f_{ 7/2}~ 0f_{ 5/2}~ 0f_{ 5/2}~ 0f_{ 7/2}$ &  5 &  0 &  0.218 \\
 $ 0f_{ 7/2}~ 0f_{ 5/2}~ 0f_{ 5/2}~ 0f_{ 5/2}$ &  5 &  0 &  0.957 \\
 $ 0f_{ 7/2}~ 0f_{ 5/2}~ 1p_{ 3/2}~ 0f_{ 7/2}$ &  5 &  0 &  0.382 \\
 $ 0f_{ 7/2}~ 1p_{ 3/2}~ 0f_{ 7/2}~ 1p_{ 3/2}$ &  5 &  0 & -1.313 \\
 $ 0f_{ 7/2}~ 1p_{ 3/2}~ 0f_{ 5/2}~ 0f_{ 7/2}$ &  5 &  0 & -0.466 \\
 $ 0f_{ 7/2}~ 1p_{ 3/2}~ 0f_{ 5/2}~ 0f_{ 5/2}$ &  5 &  0 &  0.056 \\
 $ 0f_{ 7/2}~ 1p_{ 3/2}~ 1p_{ 3/2}~ 0f_{ 7/2}$ &  5 &  0 & -1.585 \\
 $ 0f_{ 5/2}~ 0f_{ 7/2}~ 0f_{ 5/2}~ 0f_{ 7/2}$ &  5 &  0 &  0.206 \\
 $ 0f_{ 5/2}~ 0f_{ 7/2}~ 0f_{ 5/2}~ 0f_{ 5/2}$ &  5 &  0 & -0.997 \\
 $ 0f_{ 5/2}~ 0f_{ 7/2}~ 1p_{ 3/2}~ 0f_{ 7/2}$ &  5 &  0 & -0.204 \\
 $ 0f_{ 5/2}~ 0f_{ 5/2}~ 0f_{ 5/2}~ 0f_{ 5/2}$ &  5 &  0 & -2.252 \\
 $ 0f_{ 5/2}~ 0f_{ 5/2}~ 1p_{ 3/2}~ 0f_{ 7/2}$ &  5 &  0 &  0.188 \\
 $ 1p_{ 3/2}~ 0f_{ 7/2}~ 1p_{ 3/2}~ 0f_{ 7/2}$ &  5 &  0 & -1.188 \\
 $ 0f_{ 7/2}~ 0f_{ 7/2}~ 0f_{ 7/2}~ 0f_{ 7/2}$ &  6 &  0 &  0.193 \\
 $ 0f_{ 7/2}~ 0f_{ 7/2}~ 0f_{ 7/2}~ 0f_{ 5/2}$ &  6 &  0 & -0.833 \\
 $ 0f_{ 7/2}~ 0f_{ 7/2}~ 0f_{ 5/2}~ 0f_{ 7/2}$ &  6 &  0 &  0.778 \\
 $ 0f_{ 7/2}~ 0f_{ 5/2}~ 0f_{ 7/2}~ 0f_{ 5/2}$ &  6 &  0 & -2.421 \\
 $ 0f_{ 7/2}~ 0f_{ 5/2}~ 0f_{ 5/2}~ 0f_{ 7/2}$ &  6 &  0 & -1.047 \\
 $ 0f_{ 5/2}~ 0f_{ 7/2}~ 0f_{ 5/2}~ 0f_{ 7/2}$ &  6 &  0 & -2.290 \\
 $ 0f_{ 7/2}~ 0f_{ 7/2}~ 0f_{ 7/2}~ 0f_{ 7/2}$ &  7 &  0 & -2.874 \\
\end{tabular}
\end{ruledtabular}
\end{table}

\begin{table}[H]
\caption{Shell-model SP energies (in MeV) employed in present work (see text
  for details).}
\begin{ruledtabular}
\begin{tabular}{cc}
$nlj$ & SP energies \\
\colrule
$0f_{7/2}$  & 0.0 \\
$1p_{3/2}$  & 2.7 \\
$1p_{1/2}$  & 5.5 \\
$0f_{5/2}$  & 8.5 \\
\end{tabular}
\end{ruledtabular}
\label{spetab}
\end{table}

\begin{table}[H]
\caption{Effective reduced single-neutron matrix elements of the
  electric quadrupole operator $E2$ (in ${\rm e~fm^2}$).}
\begin{ruledtabular}
\begin{tabular}{cccc}
$n_a l_a j_a ~ n_b l_b j_b $ &  $\langle a || E2 || b \rangle $ \\
\colrule
 $0f_{ 7/2}~ 0f_{ 7/2}$  &   -9.307 \\
 $0f_{ 7/2}~ 0f_{ 5/2}$  &   -4.002 \\
 $0f_{ 7/2}~ 1p_{ 3/2}$  &   -7.153 \\
 $0f_{ 5/2}~ 0f_{ 7/2}$  &    4.002 \\
 $0f_{ 5/2}~ 0f_{ 5/2}$  &   -8.456 \\
 $0f_{ 5/2}~ 1p_{ 3/2}$  &    2.882 \\
 $0f_{ 5/2}~ 1p_{ 1/2}$  &   -5.064 \\
 $1p_{ 3/2}~ 0f_{ 7/2}$  &   -7.150 \\
 $1p_{ 3/2}~ 0f_{ 5/2}$  &   -2.882 \\
 $1p_{ 3/2}~ 1p_{ 3/2}$  &   -4.711 \\
 $1p_{ 3/2}~ 1p_{ 1/2}$  &   -4.944 \\
 $1p_{ 1/2}~ 0f_{ 5/2}$  &   -5.063 \\
 $1p_{ 1/2}~ 1p_{ 3/2}$  &    4.944 \\
\end{tabular}
\end{ruledtabular}
\label{tableeffn}
\end{table}

\begin{table}[H]
\caption{Effective reduced single-proton matrix elements of the
  electric quadrupole operator $E2$ (in ${\rm e~fm^2}$).}
\begin{ruledtabular}
\begin{tabular}{cccc}
$n_a l_a j_a ~ n_b l_b j_b $ &  $\langle a || E2 || b \rangle $ \\
\colrule
 $0f_{ 7/2}~ 0f_{ 7/2}$  &  -18.241 \\
 $0f_{ 7/2}~ 0f_{ 5/2}$  &   -6.899 \\
 $0f_{ 7/2}~ 1p_{ 3/2}$  &  -18.562 \\
 $0f_{ 5/2}~ 0f_{ 7/2}$  &    6.899 \\
 $0f_{ 5/2}~ 0f_{ 5/2}$  &  -17.313 \\
 $0f_{ 5/2}~ 1p_{ 3/2}$  &    8.573 \\
 $0f_{ 5/2}~ 1p_{ 1/2}$  &  -15.934 \\
 $1p_{ 3/2}~ 0f_{ 7/2}$  &  -18.562 \\
 $1p_{ 3/2}~ 0f_{ 5/2}$  &   -8.573 \\
 $1p_{ 3/2}~ 1p_{ 3/2}$  &  -15.156 \\
 $1p_{ 3/2}~ 1p_{ 1/2}$  &  -15.579 \\
 $1p_{ 1/2}~ 0f_{ 5/2}$  &  -15.862 \\
 $1p_{ 1/2}~ 1p_{ 3/2}$  &   15.579 \\
\end{tabular}
\end{ruledtabular}
\label{tableeffp}
\end{table}

\bibliographystyle{apsrev}
\bibliography{biblio}

\end{document}